\documentclass[compsoc,conference,a4paper,10pt,times]{IEEEtran}

\usepackage{cite}
\usepackage{amsmath,amssymb,amsfonts}
\usepackage{algorithmic}
\usepackage{graphicx}
\usepackage{textcomp}
\usepackage{bmpsize}
\usepackage{xcolor}
\usepackage{lipsum}
\usepackage{subcaption}
\usepackage{booktabs} 
\usepackage{algorithm}
\usepackage{algorithmic}

\PassOptionsToPackage{hyphens}{url}
\usepackage[colorlinks=true,urlcolor=black,breaklinks=true]{hyperref}

\def\BibTeX{{\rm B\kern-.05em{\sc i\kern-.025em b}\kern-.08em
    T\kern-.1667em\lower.7ex\hbox{E}\kern-.125emX}}
\begin{document}

\author{\IEEEauthorblockN{Xinzhang Chen, Arash Shaghaghi, Jesse Laeuchli, Salil S. Kanhere}
\IEEEauthorblockA{School of Computer Science and Engineering\\
The University of New South Wales (UNSW)\\
Sydney, Australia}\\
\IEEEauthorblockN{Contact: xinzhang.chen@unsw.edu.au}
}

\title{Beyond Life: A Digital Will Solution for Posthumous Data Management}

\maketitle

\begin{abstract}
%In the digital age, managing posthumous data is an emerging challenge, often hampered by the limited practicality of technical solutions. Existing solutions are closed-source, lacking transparency, cross-platform support, and fine-grained access control. This paper presents `Beyond Life', a cross-platform digital will solution designed to securely manage and distribute digital assets posthumously. We have developed a customized Ciphertext-Policy attribute-based Encryption (CP-ABE) known as PD-CP-ABE to ensure access to will content is managed at content-level granularity efficiently at scale. Our proposed solution functions independently of service providers and is designed to bring transparency and control to users on how the will is generated, stored, and executed. We have developed and evaluated the performance and practicality of our proposed solution to ensure its real-world practicality. The system implementation is made publicly available. 

In the digital era, managing posthumous data presents a growing challenge, with current technical solutions often falling short in practicality. Existing tools are typically closed-source, lack transparency, fail to offer cross-platform support, and provide limited access control. This paper introduces `Beyond Life', a cross-platform digital will management solution designed to securely handle and distribute digital assets after death. At the core of this solution is a customized Ciphertext-Policy Attribute-Based Encryption (CP-ABE) scheme, referred to as PD-CP-ABE, which enables efficient, fine-grained control over access to will content at scale. Unlike existing systems, Beyond Life operates independently of service providers, offering users greater transparency and control over how their will is generated, stored, and executed. The system is also designed to be portable, allowing users to change their will service provider. The proposed system has been fully developed and rigorously evaluated to ensure performance and real-world feasibility. The system implementation is made publicly available. 

\end{abstract}


\section{Introduction}
We deal with a wide range of online services in the Internet era daily. From using email to connect with colleagues to sharing family photos on social media platforms, our digital presence is vast and varied \cite{Solomon2023, Kaplan2023, Hart2021, Jargon2023}. However, have you ever wondered what happens to your data during your lifetime after you pass away? You may wish to permanently erase your digital footprints, making your digital identity vanish from the Internet, exercising the 'right to be forgotten' \cite{Wolford2023, Allegri2022, Lalor2021}. You may consider applying access controls on various accounts posthumously to prevent identity
ghosting \cite{Scheldt2023, Popov2024}. You may also want to pass on your online accounts to your next of kin or friends for them to manage \cite{Doyle2023, Ignify2024}. However, most users do not have a clear digital legacy plan, and much data may be permanently lost or misused \cite{Connor2010, Shaghaghi2020}. Postmortem data management is often overlooked, and one of the main challenges is at the technological level and the adequacy of existing solutions \cite{Doyle2023, Reeves2024}. Our focus in this paper is the technological aspect of this emerging problem.

Digital will management systems can be categorized into three main types: password management tools, native platform-provided solutions, and third-party services. However, each of these approaches has notable limitations. While helpful in organizing credentials, password management tools are not designed specifically for digital estate management and typically require users to store sensitive information in a secure vault, which may not cover all digital assets. Native platform-provided tools, such as Google’s Inactive Account Manager (IAM), are limited by their lack of cross-platform compatibility. This makes it difficult for users to manage digital assets spread across multiple platforms, as these tools are not customizable to individual needs and can vary in functionality depending on the provider. Moreover, they often require users to rely on the platform’s continued existence, with the risk of data loss if the service is discontinued or the provider shuts down. While offering the potential to address some of these challenges, third-party solutions often struggle with trust issues, as users may be reluctant to rely on lesser-known providers due to privacy concerns. These services are also frequently criticized for lacking essential features, making them impractical for widespread adoption. As a result, many third-party providers have failed to gain a significant user base and have discontinued their services \cite{Reeves2024}.

This paper proposes "Beyond Life", an open-source Privacy Enhancing Technology (PET) solution for posthumous data management. The proposed solution automatically manages and distributes digital assets according to preset policies after the user's death. Using a newly designed encryption scheme, data access rights can be flexibly and efficiently controlled on the content level. The system combines blockchain technology and multi-cloud storage mechanisms to ensure transparency in the digital will system and reduce trust assumptions between entities. The proposed solution allows users to move their digital will between different service providers and generate their will content (i.e., retrieving information from various providers) locally or remotely.

Our key contributions are as follows:
\begin{itemize}
    \item \textbf{Cross-platform Support}: We design a cross-platform digital will system that supports the secure management and distribution of digital assets across multiple service providers and currently supports integration with X (Social media platform), Gmail (Email) and Google Drive (Cloud storage).

    \item \textbf{Content-level Granular Access Control for Digital Wills}: We develop a customized and novel variety of the Ciphertext-Policy Attribute-based Encryption (CP-ABE) known as PD-CP-ABE. This scheme allows for efficient and content-level fine-grained access control at scale. We have evaluated the proposed solution to ensure efficacy and scalability for real-world deployment.

    \item \textbf{Decentralized and Portable Architecture}: To eliminate the reliance on centralized third-party services and ensure full transparency of any action on a digital will, our system leverages blockchain as an immutable ledger for managing digital wills. The proposed system is designed to allow portability between third-party digital will service providers for a user with ease.
\end{itemize}

The remainder of this paper is organized as follows: Section \ref{Background} provides an in-depth exploration of the background and foundational concepts, detailing existing solutions in posthumous data management. Sections \ref{System Design}, \ref{Digital Will App}, and \ref{Digital Will Broker} cover the system design and architecture of "Beyond Life," including the technical specifications and interaction models. Section \ref{Threat Model} outlines the threat model for "Beyond Life," identifying potential security vulnerabilities and mitigation strategies. Section \ref{Implementation and Evaluation} discusses the implementation process and evaluates the performance of the custom encryption scheme, focusing on the improvements and comparative results. Finally, Sections \ref{Discussion} and \ref{Conclusion} summarize the key findings and highlight areas for future improvement. Additionally, Section \ref{Project Source Code} provides links to access the source code.

\section{Background}
\label{Background}
Digital assets refer to various forms of data and online resources generated or accumulated by individuals in the digital domain during their lifetime. These assets include, but are not limited to, personal emails, social media accounts, cloud storage files and digital photos \cite{Jargon2023, Connor2010, Solomon2023, Kaplan2023}. As Gulotta et al. \cite{Gulotta2017} have noted, digital assets are increasingly integral to personal and professional life, they often contain valuable information and irreplaceable memories. The management and protection of these assets, particularly after the individual's death, presents unique challenges due to their sensitive and highly private nature.

Posthumous Data Management Solutions are systems and processes designed to handle an individual's digital assets after death, ensuring they are managed according to their wishes. We categorize these solutions into three main categories:

    \textbf{Password management tools:} Although not explicitly designed for posthumous data management, password managers such as 1Password \cite{Summers2021} and LastPass \cite{Corbett2022} can serve this purpose. These applications store and manage passwords for various online accounts, and users can share their master password with a trusted individual who can then access the accounts as needed. This method relies on manual processes, which may lead to risks such as data loss or misuse if the credentials are forgotten or mishandled. More importantly, this type of solution is an all-or-nothing solution, where users provide full access to the account or nothing at all, thus unable to provide fine-grained control.

    \textbf{Native Platform-Provided tools:} Some online platforms provide built-in features to help users plan their posthumous data. For example, Facebook allows users to set up legacy contacts to memorialize their accounts and manage their profile pages after they pass away \cite{Babbar2018, TrustedWill, Popov2024}.  Legacy contacts can download copies of content shared on Facebook, update their profiles and even request to delete the account \cite{Jargon2023, FacebookHelp}. However, Instagram, another social media app under Meta, only allows accounts to be memorialized, and legacy contacts cannot be configured \cite{Jargon2023, Hartung2022, InstagramHelp}. Moreover, Google provides IAM, where users can set how to handle their data when their accounts are inactive for some time, such as setting their emails, photos, etc., to be sent to specific contacts and even deleting the account \cite{Popov2024, Jargon2023, Babbar2018, GoogleSupport}. These solutions are integrated into the platform’s ecosystem, and while providing a more seamless approach to data management, they lack practicality for users. A regular user has around 100 online accounts throughout their life \cite{CNN2024}. Even if they only want to plan their posthumous data for ten of them, they need to first understand and familiarize themselves with the specific rules of each platform and set corresponding policies for each platform independently, which is costly to operate and learn.
    
    \textbf{Third-party solutions:} The most popular solution types offer an alternative way to manage posthumous data by providing services across multiple platforms and accounts \cite{Reeves2024}. They typically allow users to specify how their digital assets are handled after death, including the distribution of passwords, social media accounts, and data stored across various platforms. For example, Perpetu\cite{Woollaston2013} is a service that can manage data across multiple platforms during one’s lifetime. Users can predefine actions for their online accounts, including Facebook, Twitter, Gmail, and GitHub, which are triggered by an official death report. Upon activation, Perpetu either sends the data to designated heirs via email or deletes the relevant content. However, Perpetu’s reliance on centralized services presents significant drawbacks. Users must trust the platform with sensitive information, which raises concerns about privacy and security. Additionally, the stability of such platforms is not guaranteed, as seen with Perpetu, which has since suspended its services, rendering its users' data inaccessible\cite{Perpetu}. Similarly, Locasto, Massimi and DePasquale \cite{Locasto2011} have proposed a potential solution for digital identity management and inheritance based on modern browser and cloud storage technologies. Their system attempts to unify the management of digital identities across different platforms by intervening in the process of creating credentials for new websites or applications. Their solution proposed the concept of identity containers, and users can transfer digital identities by transferring the corresponding container. However, this approach shares the same vulnerabilities as other centralized solutions, such as dependency on cloud providers’ security and privacy, which can be problematic for users handling sensitive data. In contrast, Shah et al.\cite{Shah2020} proposed using blockchain technology to enable the postmortem inheritance of cryptocurrencies. The proposed solution maps the digital assets using ERC-20\cite{Nico2024} and ERC721\cite{Entriken2018} tokens and enables the owner of the Will to deploy a smart contract mentioning his wishes. However, the system is narrowly focused on cryptocurrency, limiting its applicability to a broader range of digital assets.

In summary, although password management tools are straightforward,  they \textbf{lack the granularity needed for controlled access} and rely on manual processes, posing risks of misuse or data loss. Native platform-provided tools, while integrated and user-friendly within their respective ecosystems are limited by their platform-specific nature and \textbf{require considerable effort to manage multiple accounts across different services}. Although more comprehensive in managing assets across platforms, third-party solutions present serious \textbf{concerns regarding privacy, security, and the potential instability of the service providers}.

\section{System Design}
\label{System Design}
In the following sections, we will describe how the system was designed and how it meets the requirements analyzed in Section \ref{Background}. For clarity, in the subsequent description, we assume that a user currently wishes to create a digital will and has named two heirs. The user will include three digital platforms: email, cloud, and social media. In practice, our system can easily scale to handle multiple requests simultaneously when deployed on a server, and the will creator can extend coverage to additional online platforms by providing more information for the will.

\begin{figure}[h]
  \centering
  \includegraphics[width=0.6\columnwidth, trim={0cm 0cm 0cm 0cm}, clip]{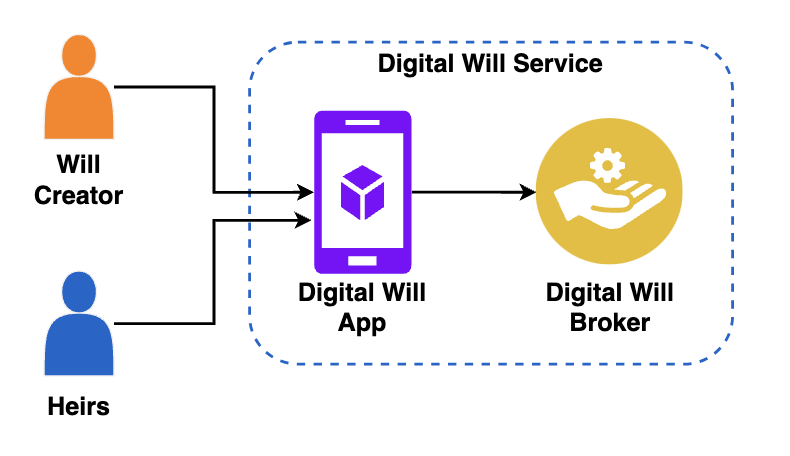}
  \caption{Main entities of the system}
  \label{System Overview}
\end{figure}

\begin{table*}[htbp]
\caption{System Key Entities}
\begin{center}
\begin{tabular}{|c|p{0.75\textwidth}|}
\hline
\textbf{Entity} & \multicolumn{1}{c|}{\textbf{Description}} \\  
\hline
Will Creator & The individual who initiates and manages the digital will. This person holds the highest authority over their digital assets and determines how they will be handled posthumously. \\
\hline
Will Heir & The designated person who will receive access to the digital assets or specific permissions upon activating the digital will. They are granted access keys based on the will creator's instructions. \\
\hline
Digital Will App & A user-friendly application provided by a digital will service provider to connect users and provide the necessary services. The will creators and heirs will use the same application to interact with the will. \\
\hline
Digital Will Broker & The entity responsible for processing digital wills, executing the will upon activation, retrieving data, encrypting it, and distributing access keys to the designated heirs. The broker also ensures the will's conditions are met in a decentralized manner. \\
\hline
\end{tabular}
\label{tab1}
\end{center}
\label{System Key Entities}
\end{table*}

\subsection{Overview}
Fig.\ref{System Overview} provides an overview of the proposed solution, illustrating how users will create a digital will using the Digital Will App \textbf{(DWA)}. Users can specify instructions for managing their digital assets through the app and designate their heirs. The app will aggregate the user's input and generate a digital will file, which is then transmitted to the broker. Upon transmission of the will to the Digital Will Broker \textbf{(DWB)}, the broker parses the digital will file, deploys it on the private blockchain as a smart contract and waits for it to be triggered. Once the specified trigger conditions are satisfied, the broker assumes the role of executor for the digital will. In this capacity, the broker retrieves the relevant data from different service providers, encrypts it, generates the appropriate access keys for the heirs, and subsequently disseminates both the keys and the encrypted data to the heirs. Table \ref{System Key Entities} lists the key entities of the system and describes their main role. We detail the specific components of the proposed solution in the following sections.

\subsection{Design Considerations}
\label{Design considerations}
In our system design considerations, we identified two primary interaction modes based on user survey results and technical system requirements. 

The first mode caters to scenarios where the digital will creator may prefer that the broker not access their sensitive information when generating the will content. In this mode, the DWA runs on the user's devices to handle data retrieval, encryption, splitting, and the creation of the will locally on their device. Hence, by keeping the processing of sensitive data entirely on the creator's local device, the broker's role is limited to receiving and deploying the generated will and facilitating interactions for will activation with heirs. This approach may require significant computational and storage capabilities on the creator's local device, making this impractical for most everyday users. 

The second interaction mode involves the will creator using the DWA primarily to generate the digital will. In this mode, the creator provides access tokens for data retrieval from various platforms, defines access policies for different content, and specifies relevant information about the heirs within the app. The app then packages these details into a digital will and sends it to the broker. Upon successful will activation, the broker uses the provided access tokens to retrieve data and performs encryption, splitting, and distributing the data on the broker server. In this case, the will creator must trust the broker, as the broker can access all the data during the retrieval. However, our system mitigates this risk by including an independent Broker Behavior inspector. Every action the broker performs is recorded on a private blockchain, allowing other platform participants to verify the correctness of the broker's actions and enhancing transparency. The Broker Behavior inspector continuously monitors the broker's actions and responds immediately if any anomalies are detected.

In the following descriptions of DWA and DWB, we will only address the second mode because this mode is more similar to the previously proposed solution, which relies on third-party servers to do most of the computations rather than local devices. Due to the limitations of local resources and the monitoring mechanisms mentioned above, which can help reduce the risk of broker misuse to a certain extent, the second mode of interaction will likely be the primary approach.

\section{Digital Will App}
\label{Digital Will App}
The DWA is a software application provided by the digital will service provider. It will let the will creator and the heirs interact with the DWB using their local smartphones or PCs. The primary function of the DWA is to serve as a tool for will creators to generate digital wills. It needs to capture the preferences of the will creators, generate the corresponding will documents, and transmit these documents to the broker for deployment. The will creators can connect their accounts from various online platforms through the interfaces and get the corresponding access token. These tokens will be used to retrieve data from different platforms' servers and display the data to the creator. The creator can then customize access policies for specific content to restrict access for heirs. Additionally, the creator must input relevant information about the heirs and the corresponding access authorization parameters through the DWA. We assume the creator and the heirs must have pre-registered accounts on DWA.

\subsection{Overview}

The DWA primarily requires the will creator to provide the following information to generate a digital will. First, the creator must authorize the DWA to connect to their other online platforms, allowing the DWA to get the corresponding access tokens. Then, the DWA will use these tokens to retrieve the data and display it to the creator. The creator can assign appropriate access policies to the data based on its content through the provided application interface. For instance, the creator can restrict social media posts related to family gatherings so that only family members can access them.

Once the access policies are set, the creator must designate heirs through the DWA, who will receive the corresponding access permissions upon the will's activation. Here, it is assumed that the heirs must have pre-registered accounts through the DWA, enabling the subsequent key generation module to create keys based on the users' IDs. Finally, the will creator must establish the conditions for triggering their digital will. This can be set to a threshold-based trigger, where the activation of the will requires a voting mechanism. The number of votes favouring activating the will must meet or exceed the predetermined threshold for the will to be triggered. Alternatively, the creator can allow third-party institutions to intervene in the triggering process to reduce the risk of abnormal will activation. Please refer to Section \ref{Digital Will Trigger Module} for detailed descriptions.

\subsection{Digital Will Generation Module}
\begin{figure}[h]
\centerline{\includegraphics[width=0.75\columnwidth, trim={0cm 0cm 0cm 0cm}, clip]{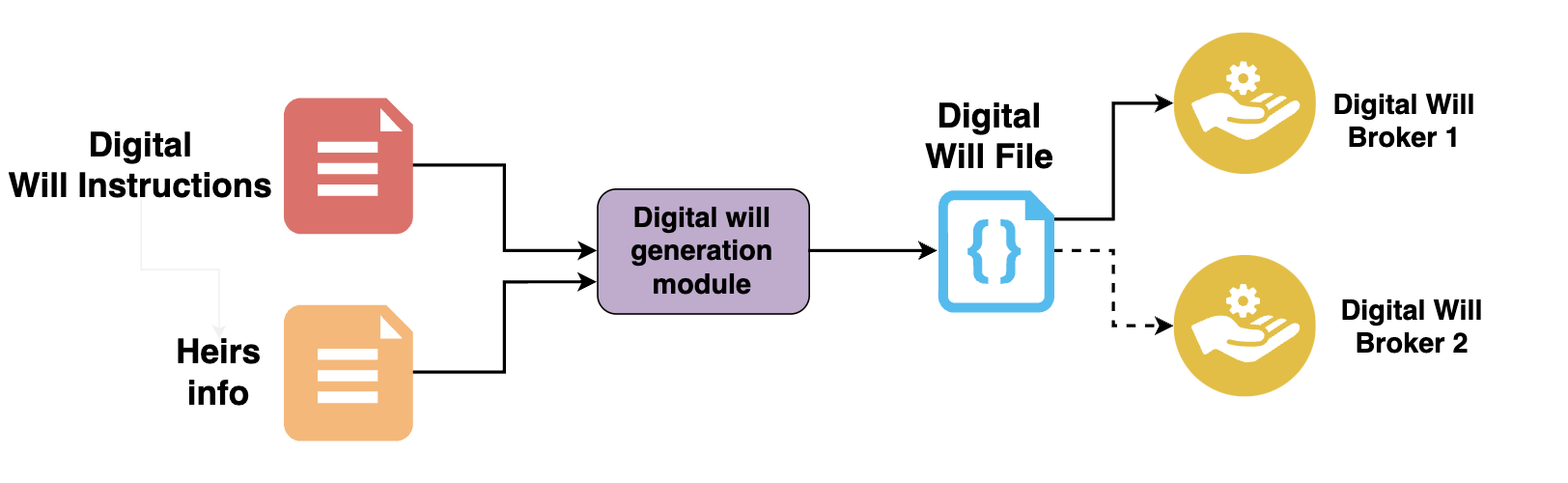}}
  \caption{Digital Will Generation Module}
  \label{Digital Will Generation Module}
\end{figure}

As shown in Fig.\ref{Digital Will Generation Module}, the Digital Will Generation Module is responsible for creating digital wills based on the inputs the creator provides through the DWA. This module processes two main inputs: Digital Will Instructions and Heirs' Information.

\begin{itemize}
    \item Digital Will Instructions: these are the instructions provided by the Will Creator, detailing how their digital assets should be managed and distributed upon death. This includes specifying which assets should be deleted or transferred. Also, the Will Creator can specify how the will will be triggered, changing the effect of the Digital Will Trigger Module.

    \item Heirs' Information: This includes the details of the individuals designated as heirs by the Will Creator. It encompasses their identities, contact information, and the specific access rights they are entitled to access.
\end{itemize}

The Digital Will Generation Module consolidates these inputs to produce a Digital Will File. This file encapsulates all the instructions and the heirs' details in a structured format, which the DWB can securely process.

 As pointed out by Allison et al.\cite{Allison2023}, existing technical solutions often fail to operate in the long term due to the lack of a suitable business model; therefore, system portability is a critical aspect to consider for the broad applicability and interoperability of the solution. In our system, we propose to use XML as the digital will file format to enable interaction and synchronization between different service providers. The file format includes detailed metadata, such as information about assets, heirs and access control policies in its hierarchical structure. Different service providers can also extend this file format, adding additional metadata or features to meet their specific requirements and business logic without affecting the core functionality of the digital will system. Notably, the usability of the digital will file does not depend on the continued operation of any single service provider. In cases where a service provider ceases its operations, other providers can parse and redeploy the previously created wills as long as they adhere to the XML schema.

 The Digital Will File is then handed over to one or more Digital Will Brokers (as shown by the dual arrows in Fig.\ref{Digital Will Generation Module}). These brokers ensure the secure execution and storage of the digital will, ready to be triggered upon verification of the Will Creator's death.

\begin{figure}[h]
\centerline{\includegraphics[width=0.75\columnwidth, trim={0cm 0cm 0cm 0cm}, clip]{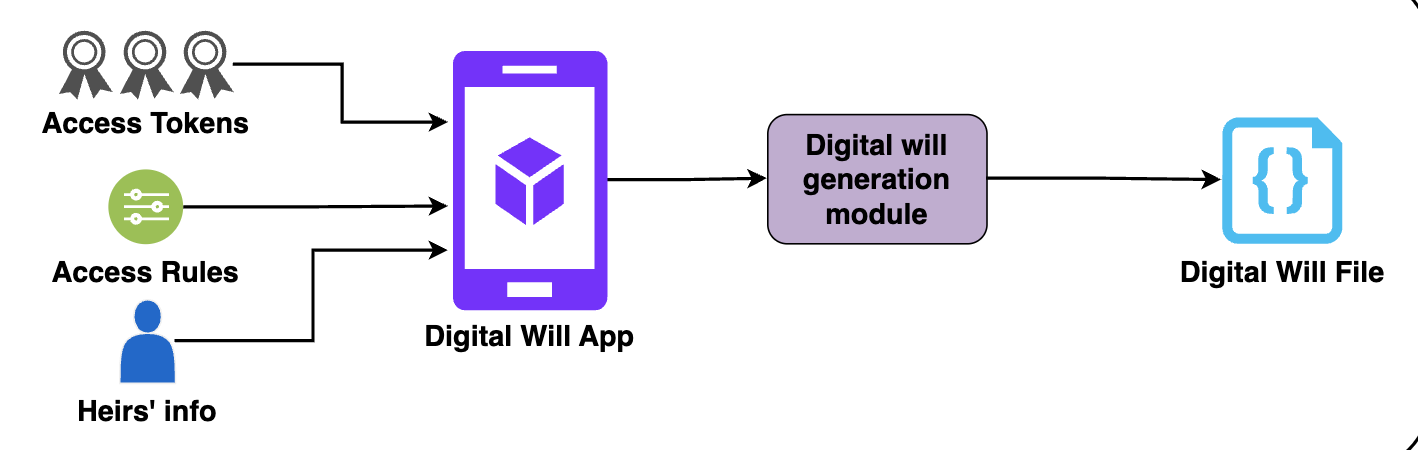}}
  \caption{DWA as a Digital Will File Generator}
  \label{DWA as a Digital Will File Generator}
\end{figure}

\begin{figure}[h]
\centerline{\includegraphics[width=0.75\columnwidth, trim={0cm 0cm 0cm 0cm}, clip]{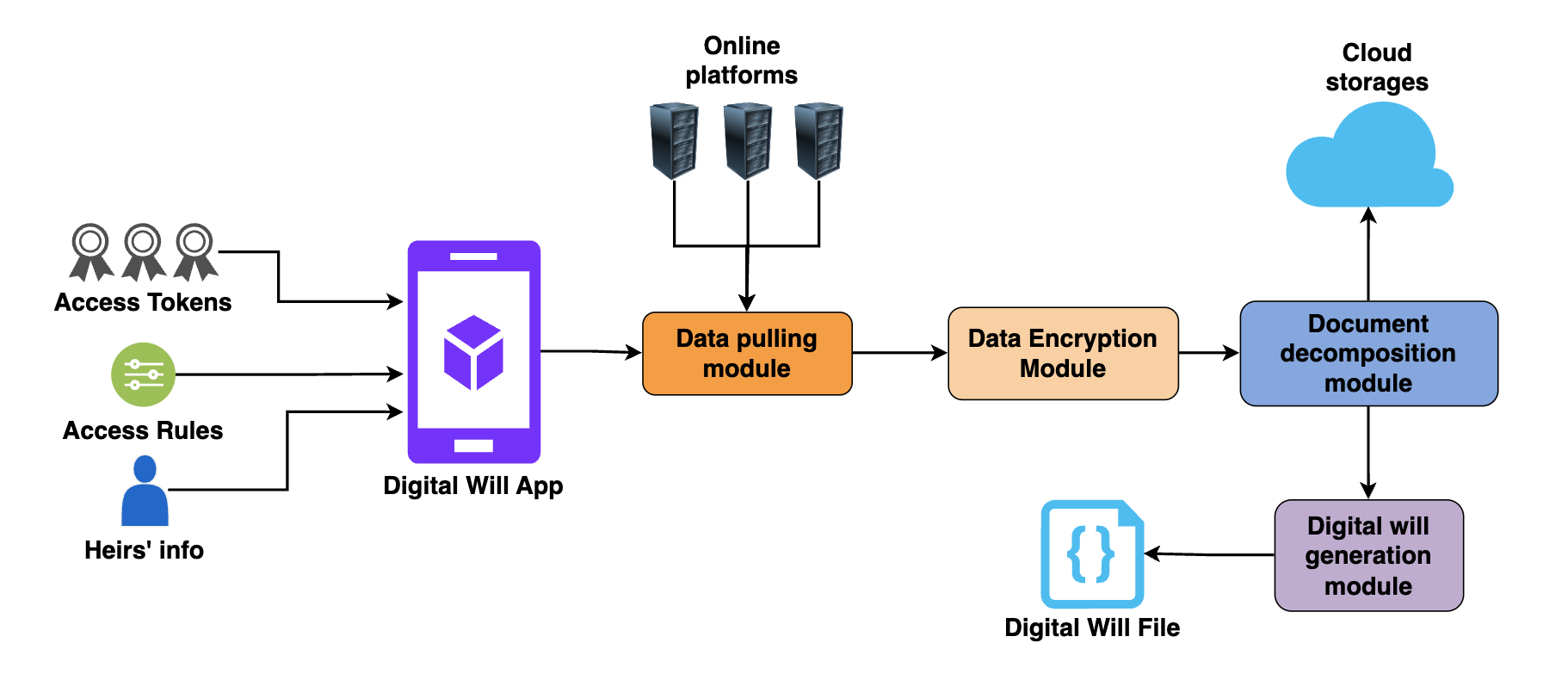}}
  \caption{DWA as an Independent Digital Will Generator}
  \label{DWA as an Independent Digital Will Generator}
\end{figure}

As shown in Fig.\ref{DWA as a Digital Will File Generator} and Fig.\ref{DWA as an Independent Digital Will Generator}, depending on the creator's concerns about data privacy, the DWA can either serve as a tool for generating the digital will file or act as a comprehensive digital will generator. In the latter case, the DWA would undertake most of the functions typically performed by the broker, effectively assuming the broker's role in the digital will creation process. For Fig.\ref{DWA as an Independent Digital Will Generator}, we outline the sequence of interactions for each module, with detailed descriptions of the modules provided in Section \ref{Digital Will Broker}, which discusses the DWB.

\section{Digital Will Broker}
\label{Digital Will Broker}
\begin{figure}[h]
  \centerline{\includegraphics[width=0.8\columnwidth]{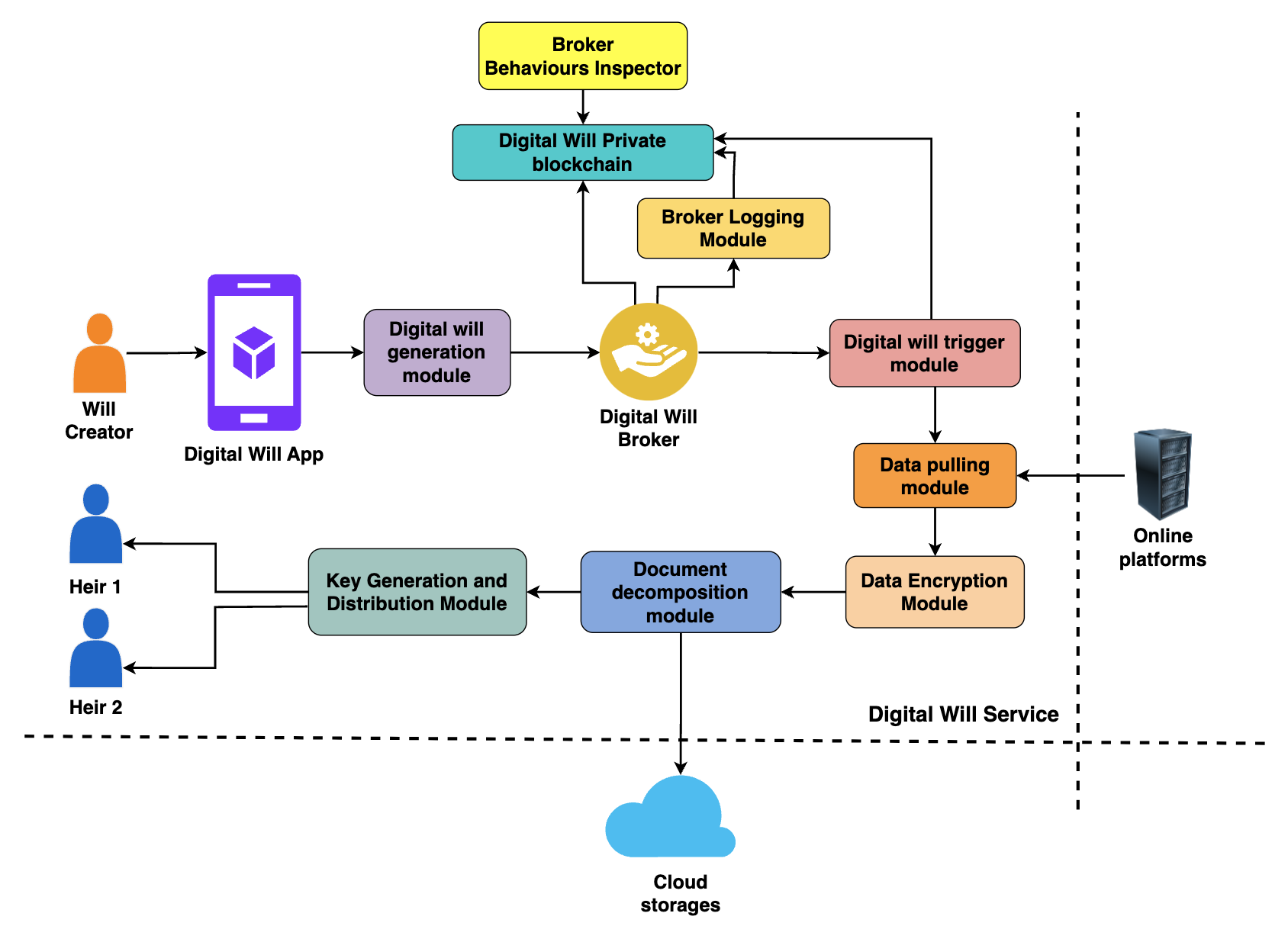}}
  \caption{Digital Will Broker Overview}
  \label{Digital Will Broker Overview}
\end{figure}

The DWB is a crucial component in the digital will ecosystem, acting as an intermediary between the will creators, heirs, and the digital assets stored across various platforms. Upon receiving the digital will from the creator, the DWB generates and deploys a smart contract on a private blockchain, encoding the essential parameters of the will, such as the heir's identities, access control policies, and activation conditions. Once deployed, the smart contract passively enforces these parameters, awaiting the specified conditions to be met for will activation. As shown in Fig.\ref{Digital Will Broker Overview}, once the conditions are met, the DWB actively manages the execution of the digital will according to the instructions specified by the creator. This includes the secure handling of access tokens, enforcing access policies by the encryption scheme, and distributing digital assets to heirs upon activating the will.

The DWB system architecture is designed to be secure and reliable, incorporating multiple modules to ensure the integrity and confidentiality of the process. Key modules include the \textbf{Data Pulling Module}, responsible for retrieving digital assets using access tokens; the \textbf{Data Encryption Module}, which encrypts assets to ensure secure storage and transmission; and the \textbf{Broker Logging Module}, which records all actions taken by the broker to a private blockchain, providing transparency and accountability.

Additionally, we have designed an independent \textbf{Broker Behavior Inspector module} to ensure the integrity and reliability of the DWB's operations. This module continuously monitors the broker's real-time operations and transaction records on the private blockchain. If suspicious or unauthorized actions are detected, the Inspector module can stop the actions and promptly alert the system administrators and stakeholders to take corrective measures.

The following sections will be divided into subsections based on the different modules in the system. They will first describe the specific requirements and design of the module and then analyze the actual use case.

\subsection{Digital Will Private Blockchain and Broker Monitoring System}

The \textbf{Digital Will Private Blockchain} serves as our system's main storage and logging component, ensuring the secure and immutable storage of digital wills. This blockchain operates under a consortium governance model, where multiple trusted entities (such as digital will service providers, legal entities, or regulatory bodies) act as validating nodes. It ensures that all operations related to the digital will, including creation, modification, and activation, are transparently recorded and verifiable. The immutable nature of the blockchain guarantees that once a digital will is recorded, it cannot be altered or deleted without detection, ensuring that the will creator’s instructions will be executed accurately as intended.

Along with the Digital Will Private Blockchain, the system includes two crucial modules: \textbf{the Broker Logging Module} and \textbf{the Broker Behaviours Inspector}. Together, these components form a comprehensive monitoring and auditing mechanism that ensures the integrity and transparency of the digital will management process.

The Broker Logging Module maintains a detailed and immutable log of all activities carried out by the DWB. From the initial creation and storage of digital wills to their eventual activation and execution, each action is recorded with timestamps and relevant details. In the event of any disputes or queries, the detailed logs serve as a reliable reference to trace the sequence of actions and validate the authenticity of the operations.

To further strengthen the system's security, the Broker Behaviours Inspector continuously monitors the actions of the DWB. This module operates as an external trusted entity, which may be a government-appointed auditor or an independent regulatory body, tasked with ensuring that the broker adheres to expected behavior and security policies. "Expected behaviour" in this context refers to the predefined set of actions and protocols the broker is authorized to execute, including properly handling digital wills, secure data encryption, and appropriate key distribution. The Inspector performs real-time analysis of the broker's actions, checking for deviations from these expected behaviours. If suspicious or unauthorized activities are detected, the module promptly stops the actions and logs the incidents for further investigation.

\begin{figure}[h]
  \centerline{\includegraphics[width=0.75\columnwidth, trim={0cm 0cm 0cm 0cm}, clip]{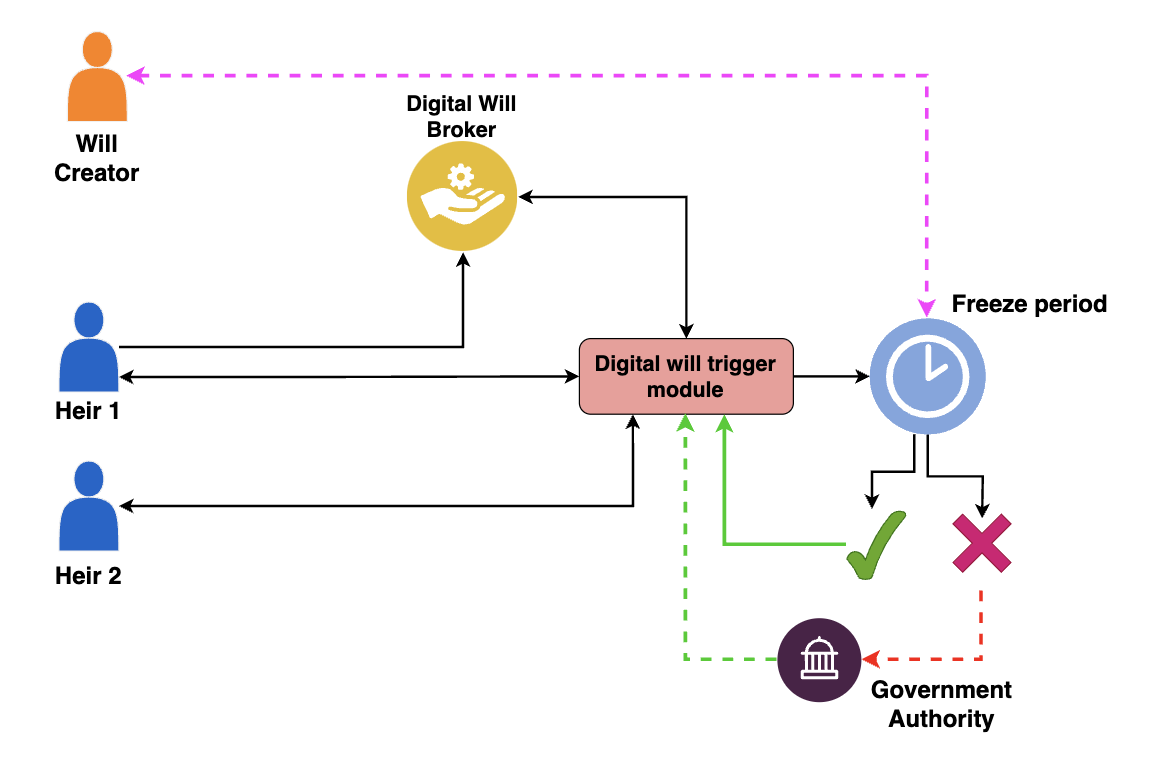}}
  \caption{Digital Will Trigger module}
  \label{Digital Will Trigger module Figure}
\end{figure}

\subsection{Digital Will Trigger Module}
\label{Digital Will Trigger Module}
The Digital Will Trigger Module is a key component that manages the activation process of the digital will, typically initiated by the heirs (see Fig.\ref{Digital Will Trigger module Figure}). Upon receiving the trigger request from the heir, the broker will update the status of the corresponding will’s smart contract to ensure that the activation request has sufficient support. The voting mechanism and activation conditions are encapsulated and encoded in the smart contract of the will, hence arbitrary modifications are not possible.

This voting mechanism involves setting specific thresholds that must be met for the will to be triggered. For instance, a minimum number of heir approvals might be required to validate the activation request. Once the voting process is initiated, the system will impose a freeze period. This period provides the will owner adequate time to contest or prevent malicious attempts to trigger the will prematurely.

Throughout the triggering process, the voting mechanism collects approvals from the eligible heirs via DWA and meticulously records and updates the status of the will. Each heir's approval is securely logged, and these status updates are stored on the private blockchain, ensuring their integrity and immutability.

The Digital Will Trigger Module will notify the DWB if the trigger conditions are met, implying that the necessary approvals are obtained. The broker then executes the following steps outlined in the digital will. Conversely, if the trigger conditions are not satisfied, heirs can seek intervention from governmental authorities. In such cases, authorities may override the system by providing official documentation, such as a death certificate or court order, that validates the triggering of the will.

\subsection{Data pulling Module}
Upon successfully triggering the digital will, the DWB initiates the data retrieval process using the Data Pulling Module. As shown in Fig.\ref{DWA as an Independent Digital Will Generator}, it systematically uses the access tokens to make API calls to the relevant online platforms. Each API call retrieves the designated data associated with the digital assets outlined in the will.

We assume that all online platforms provide relevant API interfaces and that the returned data files are in JSON format to ensure consistency and ease of processing in subsequent steps.

\begin{figure}[h]
  \centerline{\includegraphics[width=0.75\columnwidth, trim={0cm 0cm 0cm 0cm}, clip]{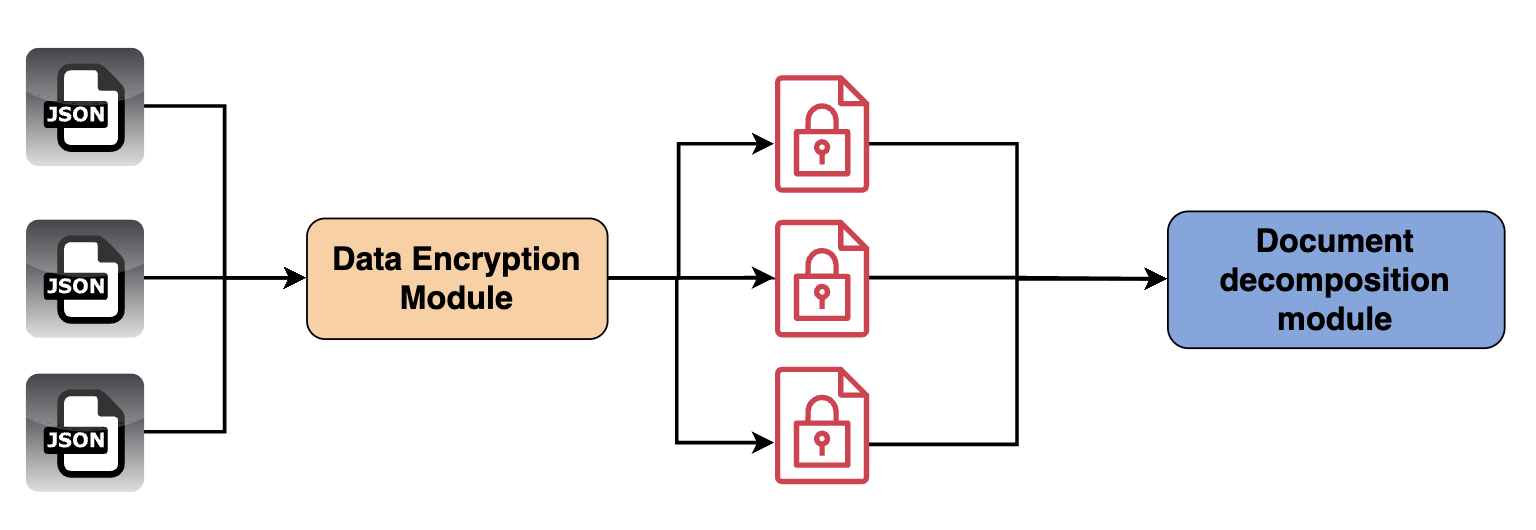}}
  \caption{Data Encryption Module with Standard CP-ABE}
  \label{Data Encryption Module with Standard CP-ABE}
\end{figure}

\begin{figure}[h]
\centerline{\includegraphics[width=0.75\columnwidth, trim={0cm 0cm 0cm 0cm}, clip]{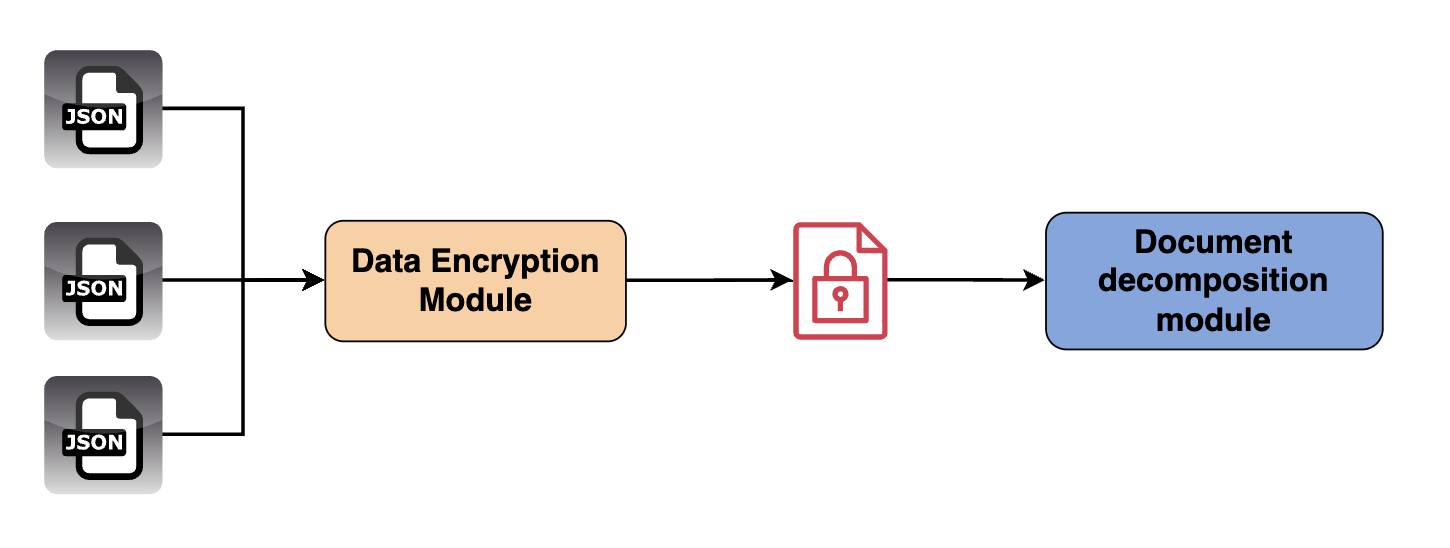}}
  \caption{Data Encryption Module with customized CP-ABE}
  \label{Data Encryption Module with customized CP-ABE}
\end{figure}

\subsection{Data Encryption Module}
The Data Encryption Module is a critical component of the DWB. Its primary function is to ensure the secure storage and transmission of the will creator's data by encrypting it before it is distributed to the designated heirs. This module is essential for preventing unauthorized access and ensuring that only authorized heirs can decrypt the data according to the will creator's specifications.

When designing this module, we consider both the efficiency of the encryption scheme and the flexibility of access control mechanisms. A typical user might have over ten online accounts included in their digital will, with data potentially comprising large files such as photos and extensive text documents. Consequently, our encryption scheme must accommodate these factors, as excessive files could impact subsequent storage processes.

Two models, role-based encryption (RBE) and attribute-based encryption (ABE), were evaluated to consider potential encryption schemes. RBE is a cryptographic model where access to encrypted data is determined by the roles assigned to users. Each role is associated with specific permissions, and users are granted decryption keys based on their current roles. While RBE is straightforward and aligns well with traditional access control mechanisms, it lacks flexibility when dealing with complex access policies that depend on multiple conditions.

ABE offers a more flexible approach by allowing access control policies to be defined based on user attributes rather than fixed roles. In ABE, both the data and the decryption keys are associated with sets of attributes. 

RBE assigns access rights based on predefined roles within an organization, which is less flexible for the dynamic and diverse nature of digital wills. In contrast, ABE allows for more granular access control by defining policies based on attributes associated with the data and users. Moreover, When considering the encryption needs of the broker, our system aims to achieve a one-to-many encryption scheme where data is encrypted once and can be decrypted by different entities using distinct keys. This approach reduces the computational load on the broker and simplifies data management compared to generating separate encrypted files for each role. Consequently, ABE is more suitable for our use case.

There are two main types of ABE: Key-Policy ABE (KP-ABE) and Ciphertext-Policy ABE (CP-ABE). In KP-ABE, the access policies are embedded in the decryption keys, and the data is encrypted with a set of attributes. A user can decrypt the data if the attributes associated with the ciphertext satisfy the policy embedded in their decryption key. This approach gives the key issuer control over the access policies. However, KP-ABE can be less flexible because modifying access policies requires reissuing keys.

In CP-ABE, the access policies are embedded in the ciphertext, and the decryption keys are associated with a set of attributes. This model lets the data owner define and modify access policies directly when encrypting the data. CP-ABE provides greater flexibility in managing access control, as policies can be updated without reissuing keys to users. Given this flexibility and the control it grants to data owners, CP-ABE is more suitable for the dynamic and attribute-rich environment of a digital will system.

To meet the requirement of content-level granularity, we need to improve the standard CP-ABE scheme to support partial decryption. Standard CP-ABE can only produce binary results in the decryption process, i.e., either successful decryption yields plain text or failed decryption yields meaningless results. In the context of digital wills, this can lead to poor space utilization and overly complex file splitting and management. As shown in Fig.\ref{Data Encryption Module with Standard CP-ABE} and Fig.\ref{Data Encryption Module with customized CP-ABE}, the standard CP-ABE scheme needs to encrypt each file individually, \begin{math}n\end{math} input files will output \begin{math}n\end{math} encrypted files, which will increase the difficulty of file management after transferring to the Document Decomposition Module. However, our encryption algorithm can encrypt multiple files uniformly, and the decryption result is no longer a success or failure; it can realise the automatic judgment of access rights through the specific information of the key. Multiple data files will output a single encrypted file.

\subsection{Document Decomposition Module}
The Document Decomposition module handles the decomposition and storage of data files. The module breaks down encrypted data files into individual fragments and stores them across multiple cloud storage providers. The will creator can choose storage preferences in the will and authorize access to the cloud storage space.

This approach enhances security by decentralizing data and reducing the risk of total data loss or unauthorized access. In the event of a breach or failure of one storage provider, data remains protected and accessible from other locations. If an attacker gains access to one of the storage providers and retrieves the files, they cannot obtain any meaningful information because each fragment is independent.

\subsection{Key Generation and Distribution Module}
The Key Generation and Distribution Module generates and distributes the private access keys necessary for accessing the encrypted digital assets. Each heir receives a key that allows them to access only the data they are authorized to view. The distribution process is conducted securely, and each key is encrypted using the heir’s public key before distribution, where the public key is obtained when the heir registers an account.

\subsection{Portability of Brokers}
During the life cycle of a digital will, the broker only acts as an intermediary to facilitate the encryption and decryption process rather than being the sole custodian of the encrypted data. If the current broker stops providing services and exports the system's initialization parameters, another broker can get the corresponding parameters. Since the data is stored across different clouds, another broker can also retrieve and reassemble the split encrypted files. Therefore, the broker has a portable architecture in design, ensuring the system's resilience and continuity.

\section{Threat Model}
\label{Threat Model}

Our system makes several critical trust assumptions regarding the entities involved. First, we assume that user devices such as personal computers and smartphones are secure and trusted environments. They are not compromised and can securely store the private access key. Next, the broker behaviour inspectors are also trusted entities, and they will continuously monitor the broker's behaviour and audit the records on the blockchain according to predefined rules. If they find any potentially dangerous behaviour, they will report it honestly. Additionally, the private blockchain used to store digital wills is another trusted component in our system. It is assumed to be immutable and tamper-proof. In contrast, the DWB is considered a semi-trusted entity responsible for processing digital wills, including encrypting data, splitting it across multiple cloud services, and distributing access keys to heirs. However, it may be attacked and not function according to the predefined rules. We treat cloud storage and the heirs as untrusted entities. We assume that cloud providers might attempt to access, modify, or deny access to the stored data. Similarly, heirs may also be able to override the triggering of a will by refusing to vote or to collude to circumvent access controls.

\section{Implementation and Evaluation}
\label{Implementation and Evaluation}
We have developed all key aspects necessary to prove the practicality of the solution proposed. In this section, we provide a detailed overview of the implementation aspects of our digital will system, focusing on the encryption scheme, the Document Decomposition and Assembly module, and the potential way to reduce trust in the server.

\subsection{The Digital Will App}

We developed a proof-of-concept prototype of DWA using React Native as the front-end framework. Leveraging the cross-platform capabilities of React Native, the application runs on iOS and Android devices. To simulate the interaction between DWA and the broker, we built a lightweight server using Expressjs. After deployment, the configured digital will is transferred from the DWA to the broker for processing.

In the current state of the application, we integrate the APIs of X (formerly Twitter), Gmail, and Google Drive, which represent the most common types of online platforms that we use daily: social media, email services, and cloud storage, respectively. For each platform, users need first to authorize DWA to access their data, and then DWA will retrieve and save the corresponding access token. We leverage the Gmail API\cite{GmailAPIQuickstart} and Google Drive API\cite{DriveAPIQuickstart} provided by Google to access user information for Gmail and Google Drive. For X, we used the latest X API v2\cite{XAPIToolsAndLibraries} to communicate with X's servers. The link to access the application source code is in Section \ref{Project Source Code}, the screenshots of the app and the QR code to run the application locally are under Appendix \ref{Appendix C}, which will require installing the "Expo Go" app on the local devices.

\subsection{Encryption scheme}
The foundation of our system security relies on a flexible encryption scheme specifically designed for the context of digital wills. This scheme is expected to address the shortcomings of the existing CP-ABE scheme that applied directly to digital wills. 

The initial concept of CP-ABE was introduced by Sahai and Waters\cite{Bethencourt2007}, where the encryption process involves setting an access policy for the data items, typically represented as a Boolean expression composed of various attributes. Decryption keys are generated based on the attribute sets of users; a user can decrypt the data only if his attributes satisfy the access policy's Boolean expression. The scheme will parse the given access policy and generate a corresponding access tree as the underlying data structure. Each leaf node will be used to represent an attribute, and each sub-node will be used to represent a threshold gate. When determining whether the user has permission to access the current data, the attribute information in the user's key will try to match the tree's leaf nodes and reach the root node from the tree's leaves through the threshold gates. If the attribute in the key meets the requirement to reach the root node, then the user has permission to access the current data, otherwise, denied access. 

While standard CP-ABE provides a flexible and fine-grained access control mechanism, it has limitations in the context of digital wills. A user may have thousands of files to include in his digital will, if we use standard CP-ABE to encrypt each file separately, although it is logically feasible, there are limitations in the performance aspect. We assume users will not set up independent and unique access policies for each file in practical cases. Different files may have the same access policy, or the access policies may share some subphrase. Moreover, after the encryption process, the same number of encrypted files will be obtained. Additionally, the latter file-splitting mechanism exacerbates this issue, leading to significant inefficiencies in storage costs. 

To address these limitations, Wang et al.\cite{Wang2014} proposed the first File Hierarchy CP-ABE (FH-CP-ABE) scheme, which encrypts multiple files using an integrated access structure. They found that in practical scenarios, the access control policies set by the same user for a group of personal data often have a hierarchical structure, such that the access structure of one file is a subset of another file's access structure. The associated attribute ciphertexts of the integrated access control tree only need to be calculated once, effectively saving computation and storage costs. However, their proposed solution was subsequently found to have security vulnerabilities in the computation of the transmission ciphertext, which could lead to users being able to access content that should not have been accessed and could not satisfy the chosen plain-text attacks (CPA) security requirements. 

Li et al.\cite{Li2009} modified the form of transmitted cipher texts, addressing the calculation between transmitted cipher texts but overlooking the issue of collusion among multiple users. Based on this cipher-text form, multiple users who decrypt sub-tree access structures can collude to decrypt the root node ciphertext. Recent work by Bai et al.\cite{Bai2024} attempts to address the aforementioned issues by introducing Collusion Resistance File Hierarchy Attribute-Based Encryption Scheme(CR-FH-CPABE), their works incorporating a Data Noise Vector and File Hierarchy with User Collusion Resistance. This approach aims to mitigate the problem of user collusion while maintaining computational efficiency within an encrypted file structure environment. Our encryption algorithm builds upon Bai et al.'s scheme, modifying it to fit the specific requirements of the digital will application scenario.

\begin{figure}[h]
  \centerline{\includegraphics[width=0.70\columnwidth, trim={0cm 0cm 0cm 0cm}, clip]{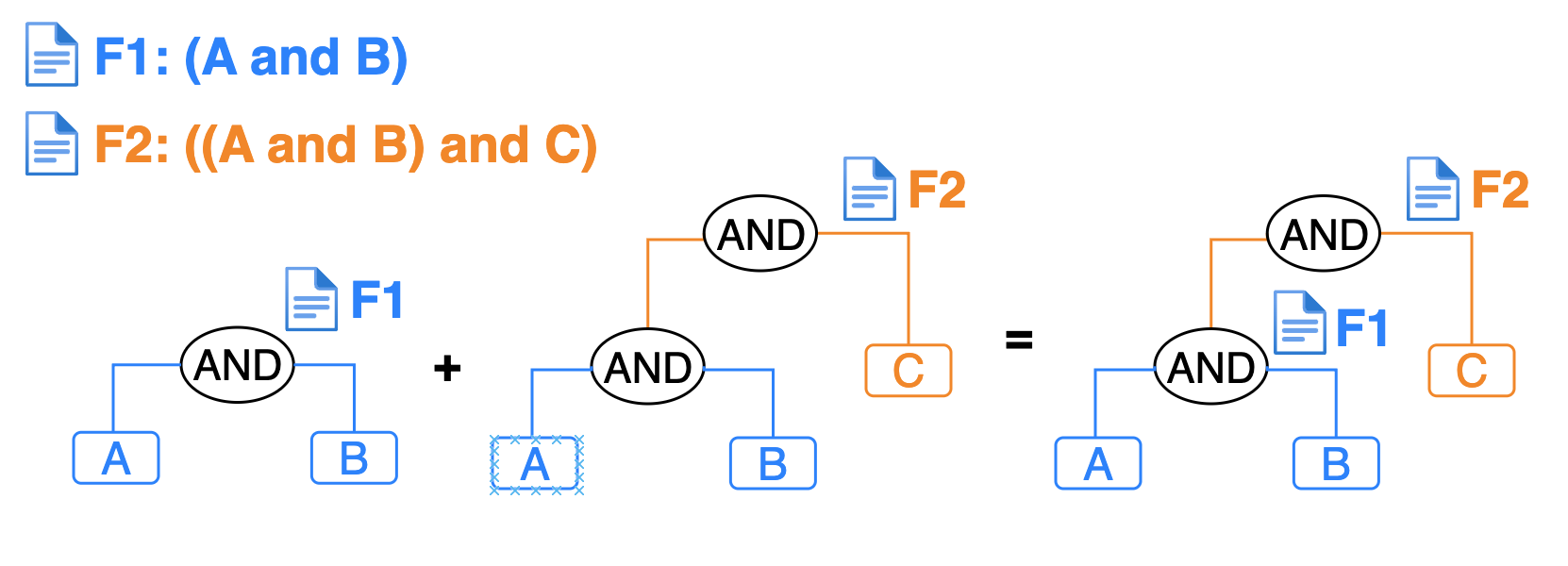}}
  \caption{Valid relationship of access structures in CR-FH-CPABE}
  \label{Valid relationship of access structures in FH-CP-ABE}
\end{figure}

\begin{figure}[h]
\centerline{\includegraphics[width=0.70\columnwidth, trim={0cm 0cm 0cm 0cm}, clip]{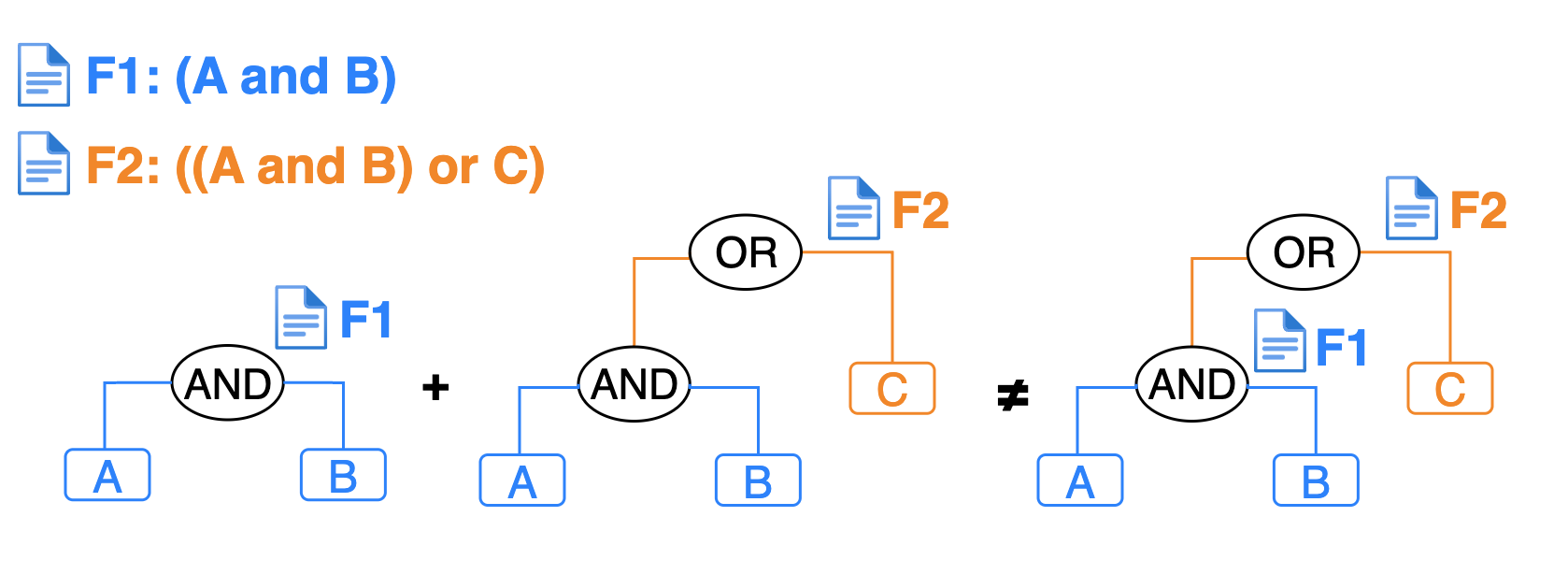}}
  \caption{Invalid relationship of access structures in CR-FH-CPABE}
    \label{Invalid relationship of access structures in FH-CP-ABE}
\end{figure}

Due to the default assumptions of CR-FH-CPABE, it cannot be directly applied to the digital will scenario. In CR-FH-CPABE, the integrated access control tree assumes a hierarchical access policy. If the current user's key satisfies one node, it can use the transmission ciphertext to directly compute all nodes below that node, thereby accelerating the decryption process. The hierarchical relationship also allows for combining multiple trees into a single tree during encryption, thus reducing computational overhead and storage consumption. As illustrated in Fig.\ref{Valid relationship of access structures in FH-CP-ABE}, if two files need to be encrypted and their access policies generate the access trees shown, we can create an integrated access tree and use it to encrypt both files simultaneously, resulting in a single encrypted file. However, if no such hierarchical relationship exists among multiple encryption objects, their access trees cannot be combined without causing errors in access policy implementation. As depicted in Fig.\ref{Invalid relationship of access structures in FH-CP-ABE}, combining these two trees improperly would lead to mistaken access control. For instance, if a user's key contains attribute \begin{math}C\end{math}, they can decrypt \begin{math}F2\end{math} and use the transmission ciphertext to decrypt \begin{math}F1\end{math}. However, \begin{math}F1\end{math} requires both attributes \begin{math}A\end{math} and \begin{math}B\end{math} for decryption, thus rendering the integrated access tree invalid. In such cases, separate access trees must be generated for each file to maintain correct access control policies.

Thus, directly using the existing CR-FH-CPABE would inevitably result in multiple encrypted files whenever there is no shared hierarchy among the encryption objects, necessitating the generation of independent trees. Even so, the scheme proposed by Bai et al. provides valuable inspiration for our design. By adopting their approach of reusing common nodes as much as possible, we can reduce computational steps during encryption and decryption, enhancing overall performance, reducing resource consumption, and preventing user collusion attacks.

\begin{figure}[h]
  \centerline{\includegraphics[width=0.65\columnwidth, trim={0cm 0cm 0cm 0cm}, clip]{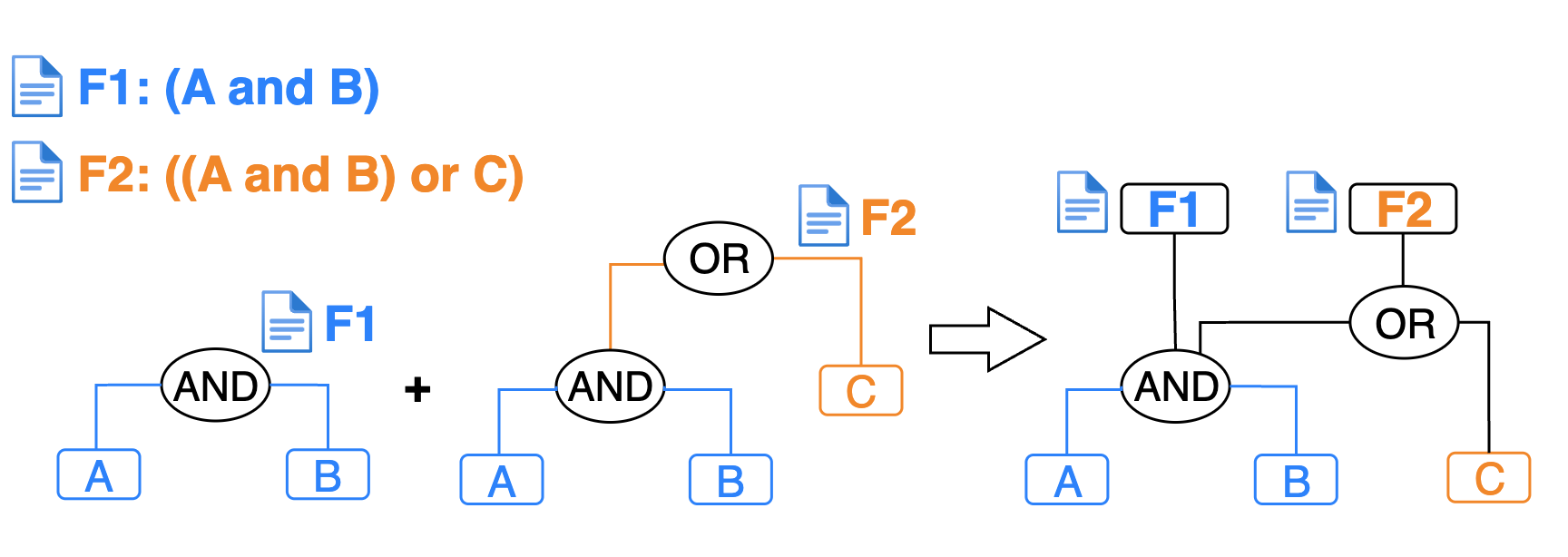}}
  \caption{Integrated access structure for PD-CP-ABE}
  \label{Integrated access structure for PD-CP-ABE}
\end{figure}

Based on this context of our requirements, we propose the Partially Decryptable Ciphertext-Policy Attribute-Based Encryption (PD-CP-ABE), an encryption scheme specifically designed for applying digital wills. For the setup and keygen functions, our scheme follows the approach proposed by Bai et al. However, we have modified the encryption and decryption algorithms and the underlying data structures.

We have redesigned the structure of the access tree, adopting a linked list-like structure to maximize node sharing and enable encryption of all objects using a single tree. 

In our scheme, users can input multiple encryption objects and a list of access policies corresponding to each object. The scheme first extracts an attribute set from all access policies and generates corresponding attribute nodes for each attribute. For the encryption objects, their respective access policies are analyzed; objects with identical policies can be grouped for encryption. We obtain a set of encryption objects and generate a data node for each object. Subsequently, the access policies for each object are parsed to connect the data nodes and attribute nodes using the link nodes. Each link node has a threshold value and two child nodes, which will be used to link the other nodes altogether. By analyzing the current policy, if an attribute node is encountered, it is directly connected to the previously generated attribute node. If a link node is encountered and it has been previously generated, it is directly connected. As shown in Fig.\ref{Integrated access structure for PD-CP-ABE}, even access policies that were previously unable to be integrated can now reuse the same nodes under the new structure.

\begin{figure}[h]
  \centerline{\includegraphics[width=0.8\columnwidth, trim={0cm 0cm 0cm 0cm}, clip]{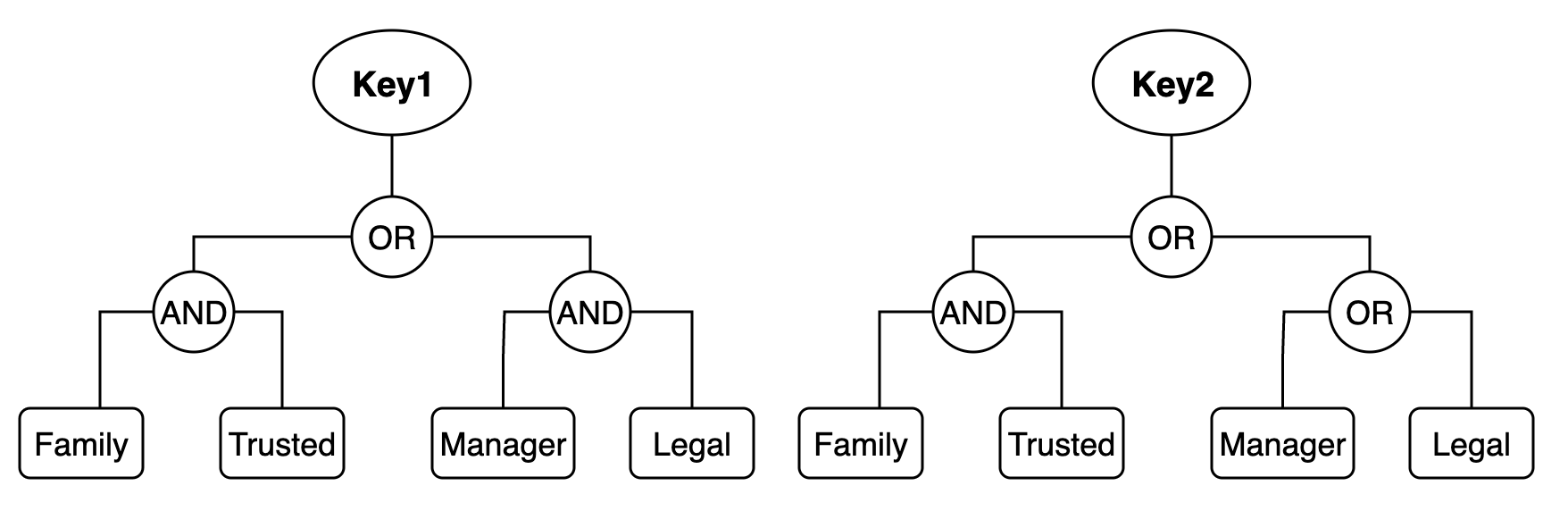}}
  \caption{Access structure for CR-FH-CPABE}
  \label{Access structure for traditional CP-ABE}
\end{figure}

\begin{figure}[h]
  \centerline{\includegraphics[width=0.50\columnwidth, trim={0cm 0cm 0cm 0cm}, clip]{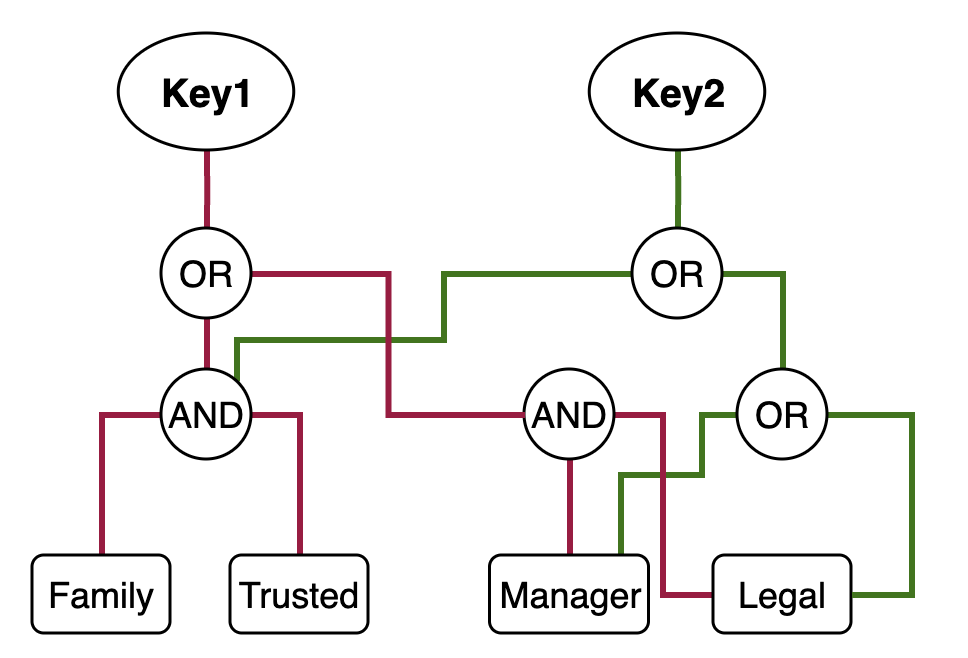}}
  \caption{Access structure for PD-CP-ABE}
  \label{Access structure for PD-CP-ABE}
\end{figure}

In our designed data structure, each node is no longer restricted to having only one parent node; it can have multiple parent nodes to reuse previously generated nodes. Furthermore, the new structure no longer has a single root node. Instead, each data node acts as an independent root node, connected to the relevant nodes as the access policy requires. Fig.\ref{Access structure for traditional CP-ABE} and Fig.\ref{Access structure for PD-CP-ABE} shows a more specific example. Fig.\ref{Access structure for traditional CP-ABE} shows two independent access structures that would be generated if CR-FH-CPABE is used. In contrast, Fig.\ref{Access structure for PD-CP-ABE} shows a new data structure that reduces the generation of duplicate nodes by connecting previously generated nodes.

During the encryption process, our scheme, similar to the one proposed by Bai et al., employs a hybrid CP-ABE method. For \begin{math}k\end{math} number of data objects, for each data object \begin{math}D_i\end{math}, a symmetric encryption key \begin{math}ck_i\end{math} is randomly selected, and AES is used to encrypt the data, and we have a set of cipher texts.

\begin{equation}
    E_{ck}(D) = \{E_{ck_1}(D_1), E_{ck_2}(D_2), \ldots, E_{ck_k}(D_k)\}. 
\end{equation}

The symmetric encryption keys \(ck = \{ck_1, ck_2, \dots, ck_k\}\) are encrypted in our scheme. We generate an integrated access tree \(T\) and assign each symmetric key \(ck_i\) to its corresponding data node \(D_i\). For each node \(N_i\), a random index \(\text{index}(N_i) \in \mathbb{Z}_p\) is selected. For each data node \(D_i\), a secret \(s_i \in \mathbb{Z}_p\) and a data noise vector \(\epsilon_i \in \mathbb{Z}_p\) are randomly chosen. We then compute the following:

% The symmetric encryption keys \begin{math}ck = \{ck_1, ck_2,...,ck_k\}\end{math} is what our scheme is going to encrypt. We generate an integrated access tree \begin{math}T\end{math} as mentioned above and assign each symmetric key \begin{math}ck_i\end{math} to its corresponding data node \begin{math}D_i\end{math}. For each data node \begin{math}D_i\end{math}, we randomly choose a secret number \begin{math}s_i \in \mathbb{Z}_p\end{math} and a data noise vector \begin{math}\epsilon_i \in \mathbb{Z}_p\end{math}, then we compute the following:

\begin{equation}
C_{D_i} = ck_i \cdot e(g,g)^{\alpha(s_i + \epsilon_i)}
\end{equation}

\begin{equation}
C'_{D_i} = f_1^{(s_i + \epsilon_i)}
\end{equation}

\begin{equation}
C''_{D_i} = f_2^{(s_i + \epsilon_i)}
\end{equation}

\begin{equation}
C'''_{D_i} = f_3^{(\epsilon_i)}
\end{equation}

Next, similar to the scheme proposed by Bai et al., we generate a polynomial \begin{math}P_{D_i}\end{math} for each data node, setting \begin{math}P_{D_i}(0) = s_i\end{math}, where \( s_i \) represents the secret associated with the data node \( D_i \). This polynomial is then propagated downwards through the tree structure. For each link node \begin{math}L_i\end{math}, we first compute its secret value \begin{math}S_{L_i}\end{math} as \begin{math}S_{L_i} = P_{\text{parent}(L_i)}(\text{index}(L_i))\end{math}, where \( \text{parent}(L_i) \) refers to the immediate parent node of \( L_i \) in the hierarchical access tree. The parent node is responsible for passing down its secret information through the polynomial function, evaluated at the index of the child link node. However, our scheme makes adjustments in the polynomial generation for link nodes to utilize the benefits of the integrated access tree.

Depending on the three scenarios, we can reuse the secret values of nodes during computation. During encryption, we maintain a hash table \begin{math}H\end{math} to record the values of nodes that have already been computed. If both the left and right child nodes of the current link node \begin{math}L_i\end{math} have values or none have values, we must recompute the secret value, generate the polynomial \begin{math}P_{L_i}\end{math}, set \begin{math}P_{L_i}(0) = s_{L_i}\end{math}, and propagate this polynomial downwards. If the threshold gate for the current link is "or," we also need to recompute the secret value as in the previous case. However, for a link node with an "and" threshold, if either the left or right child node has been previously computed, we can reuse the previously computed value as one of the inputs to generate a new polynomial.

In our scheme, when we encounter an "and" threshold link node \begin{math}L_x\end{math} with one child node already computed, we utilize the pre-computed value to help generate the polynomial for \begin{math}L_x\end{math}. Specifically, consider an "and" threshold link node \begin{math}L_x\end{math} with coordinates \begin{math}(\text{index}(L_x), s_{L_x})\end{math} and a previously computed link node \begin{math}L_y\end{math} with coordinates \begin{math}(\text{index}(L_y), s_{L_y})\end{math}. We can determine the polynomial passing through these two points using Lagrange interpolation. 

Given points \begin{math}(x_1, y_1), (x_2, y_2), \ldots, (x_n, y_n)\end{math}, the Lagrange basis polynomial \begin{math}l_i(x)\end{math} is defined as:

\begin{equation}
l_i(x) = \prod_{\substack{0 \le j < n \\ j \neq i}} \frac{x - x_j}{x_i - x_j}
\end{equation}

The Lagrange polynomial \( P(x) \) is then given by:

\begin{equation}
P(x) = \sum_{i=0}^{n-1} y_i \cdot l_i(x)
\end{equation}

In our case, we have:

\begin{equation}
L_x = (\text{index}(L_x), s_{L_x}), \quad L_y = (\text{index}(L_y), s_{L_y})
\end{equation}

Thus, the polynomial \( P(x) \) passing through these points can be computed as:

\begin{equation}
P(x) = s_{L_x} \cdot l_x(x) + s_{L_y} \cdot l_y(x)
\end{equation}

where 

\begin{equation}
l_x(x) = \frac{x - \text{index}(L_y)}{\text{index}(L_x) - \text{index}(L_y)}
\end{equation}

and 

\begin{equation}
l_y(x) = \frac{x - \text{index}(L_x)}{\text{index}(L_y) - \text{index}(L_x)}
\end{equation}

Subsequently, the polynomial is propagated to the other un-computed nodes connected to \begin{math}L_x\end{math}. For nodes that have already been computed, it is unnecessary to propagate the polynomial further. This approach not only minimizes redundant computations for already computed nodes but also reduces the bi-linear mapping calculations required for the involved attribute nodes.

When the polynomial reaches an attribute node, for each attribute node \begin{math}A_i\end{math}, it is necessary to compute the corresponding attribute cipher-texts. Unlike the scheme proposed by Bai et al., where an attribute node has exactly one pair of attribute ciphertexts, in our scheme, due to the reuse of attribute nodes, a single attribute node might have multiple pairs of attribute ciphertexts.

\begin{equation}
\left\{ \hat{C}_{x,j} = g^{q_{x,j}(0)}, \hat{C}_{x,j} = H_1(\text{att}(x))^{q_{x,j}(0)} \right\}_{j=1}^n
\end{equation}
where \begin{math}\hat{C}_{x,j}\end{math} and \begin{math}\hat{C}_{x,j}\end{math} denote the multiple pairs of attribute ciphertexts for node \begin{math}x\end{math}, indexed by \begin{math}j\end{math}.

During decryption, our scheme operates similarly to the one proposed by Bai et al. When heirs attempt to decrypt the ciphertext using their secret key \begin{math}SK\end{math}, the key's attributes are first matched with the corresponding attribute nodes. The decryption process proceeds upward through the tree, with secret values being recovered at each link node using Lagrange interpolation. 

 A key distinction of our scheme lies in the encryption process, where specific nodes utilize values from shared nodes during downward propagation. Consequently, during decryption, we also maintain a hash table \begin{math}H_2\end{math} to track all shared values that were encrypted and are now being decrypted. This approach can reduce the need for redundant bilinear map computations.

\subsection{Document decomposition and assembly}
After encrypting the files, we also need to store these files across multiple locations and make them accessible in the future. We have designed the following two algorithms to ensure document integrity and availability even in the event of partial cloud failures. Both algorithms are deployed in the document decomposition module. Due to its modularity, this module can run independently from other modules in the system, providing flexible scalability.

\subsubsection{Document splitting}

After encryption with PD-CP-ABE, each platform linked to the will creator will have a corresponding encrypted file. The broker’s next task is to partition these encrypted files according to the specifications set by the will creator. These files are then distributed across the \begin{math}N\end{math} distinct online storage locations provided by the creator. Here, \begin{math}N\end{math} refers to the total number of storage providers designated by the will creator. Each encrypted file is divided into \begin{math}N\end{math} shares using Shamir's secret-sharing scheme, where each share individually holds no meaningful information. By default, the system sets the threshold to \begin{math}N/2\end{math}, meaning that at least half of the chunks are required to reconstruct the original file. However, we also allow users to set their thresholds based on their needs. A higher threshold can offer better security, as an attacker must successfully compromise more platforms to obtain the required number of chunks, which is practically very challenging. Algorithm \ref{Splitting and Uploading Encrypted Files to Cloud} details the specifics of this algorithm, which ensures the secure and efficient distribution of the encrypted data across multiple storage locations.

\begin{algorithm}
\caption{Splitting and Uploading Encrypted Files to Cloud}
\label{alg:split_upload}
\begin{algorithmic}[1]
\REQUIRE Encrypted file $F$, number of storage locations $N$, threshold $T$
\ENSURE Shares $S_i$ stored across storage locations $C_i$
\STATE Generate $N$ shares $S_1, S_2, \ldots, S_N$ from file $F$ using Shamir's secret sharing scheme with threshold $T$
\FOR{each share $S_i$, where $i = 1, 2, \ldots, N$}
    \STATE Label $S_i$ with corresponding File ID and Share ID
    \STATE Upload share $S_i$ to cloud storage location $C_i$
    \STATE Record the mapping [File ID, Share ID, $C_i$] in the digital will's smart contract
\ENDFOR
\end{algorithmic}
\label{Splitting and Uploading Encrypted Files to Cloud}
\end{algorithm}

\subsubsection{Document Aggregation for Retrieval}
Upon receiving a data retrieval request from the heir, the broker initiates the recovery process by referencing the smart contract associated with the digital will. This smart contract contains the necessary information about the locations and IDs of the encrypted shares stored across the different cloud providers. The broker then attempts to pull the data from various storage locations, and the broker sends requests to only the minimum number of servers required to satisfy the threshold initially, thereby reducing bandwidth consumption. If some storage servers are unresponsive or encounter issues, requests can be sent to additional servers. Since we employ Shamir’s secret sharing scheme, only a threshold number of chunks are needed to reconstruct the original data successfully. Once the broker obtains the necessary chunks, it uses Shamir’s secret sharing scheme to recover the original information and return it to the heir. Algorithm \ref{Combining File from Cloud Storage} provides a pseudo-code representation of this process.

\begin{algorithm}
\caption{Combining File from Cloud Storage}
\label{alg:combine_file}
\begin{algorithmic}[1]
\REQUIRE Storage locations $C_1, C_2, \ldots, C_N$, threshold $T$
\ENSURE Reconstructed file $F$
\STATE Retrieve share information (File ID, Share ID, $C_i$) from the digital will's smart contract
\STATE Request shares $S_i$ from storage locations $C_i$ as specified by the smart contract
\STATE Verify that at least $T$ shares are successfully retrieved
\IF{number of retrieved shares $\geq T$}
    \STATE Apply Shamir's secret sharing reconstruction algorithm to the shares $S_i$ to recover the original file $F$
\ELSE
    \STATE \textbf{Error:} Insufficient shares to reconstruct the file
\ENDIF
\end{algorithmic}
\label{Combining File from Cloud Storage}
\end{algorithm}

\subsection{Reduce trust in the server}

In our system, despite mechanisms regulating broker behaviour, users must still trust the broker, primarily since both key access and encryption operations are performed at the broker's end. This setup could potentially allow a dishonest broker to access data unauthorizedly. We propose an additional layer of public key encryption to enhance trust and extend our system's security architecture. In this extended scheme, ciphertexts encrypted using PD-CP-ABE are further encrypted with the heirs' public keys.

In the scenario where the will creator primarily uses the Digital Will App to generate the digital will and lets the broker do the data pulling, we assume that the broker may access the complete data once but cannot decipher it independently once it has been distributed. This approach ensures the broker cannot interpret the data alone after its distribution. However, a drawback of this method is that it generates a unique encrypted version for each user, which could increase the complexity of managing encrypted files.

\begin{figure}[h]
  \centerline{\includegraphics[width=0.7\columnwidth, trim={0cm 0cm 0cm 0cm}, clip]{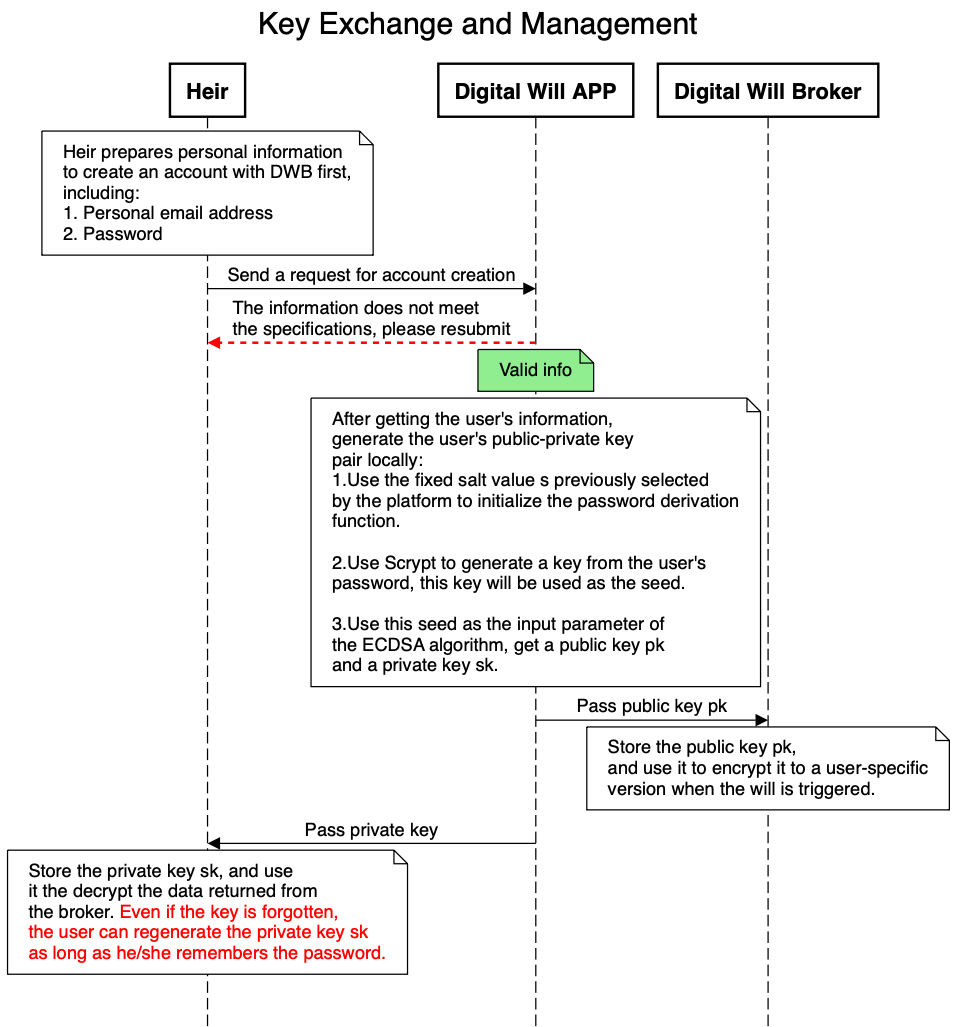}}
  \caption{Key Exchange and Management mechanism}
  \label{Key Exchange and Management mechanism}
\end{figure}

Fig.\ref{Key Exchange and Management mechanism} illustrates the key management mechanism in our system. Given that all heirs are required to create an account on the platform, the corresponding public-private key pairs are generated at that time. Our system employs a Key Derivation Function (KDF) to minimize the involvement of third-party key management services. The mechanism operates as follows:

\begin{enumerate}
    \item When an heir creates an account, the Digital Will App locally initializes the KDF function using a system-default salt.

    \item The heir's provided password is used as input to the KDF function, resulting in the generation of the private key \begin{math}sk\end{math}.

    \item The public key \begin{math}pk\end{math} is generated using the ECDSA algorithm with \begin{math}sk\end{math}.

    \item The Digital Will App returns the private key \begin{math}sk\end{math} to the heir while transmitting the public key \begin{math}pk\end{math} to the broker.

    \item Upon activation of the corresponding digital will, the broker encrypts the data for the heir using the associated public key.

    \item The heir can decrypt the data using the private key. Even if the heir loses the private key, it can be recovered if the heir remembers the password.
\end{enumerate}

\subsection{Evaluation}
We implemented our encryption scheme using the \texttt{charm-crypto} library in Python. The key components of this project have been made open source and accessible via the links listed in Section \ref{Project Source Code}.

\begin{figure}[h]
  \centerline{\includegraphics[width=1.2\columnwidth, trim={0cm 0cm 0cm 0cm}, clip]{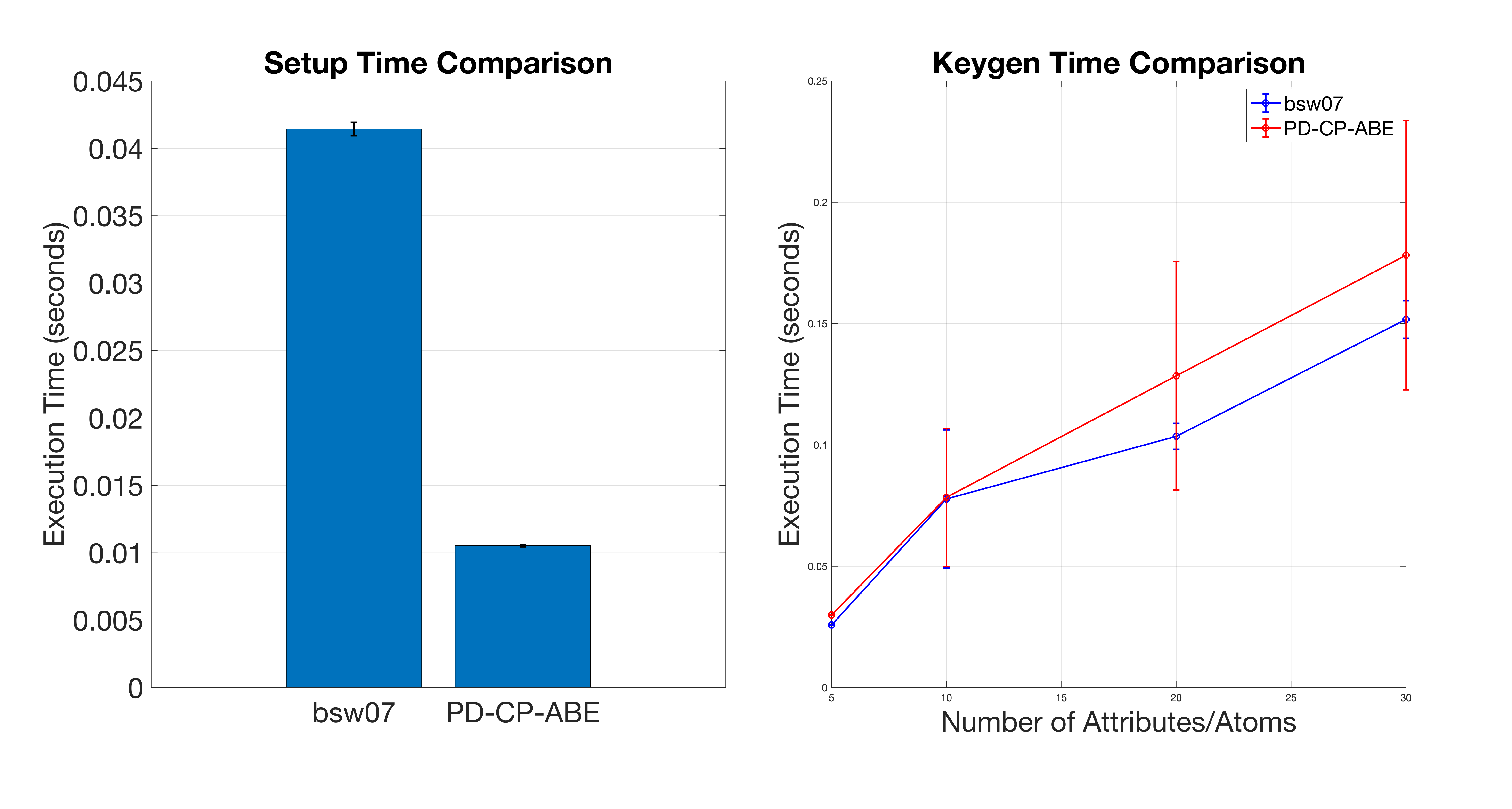}}
  \caption{Setup Time and Keygen Time Comparison for bsw07 and PD-CP-ABE schemes.}
  \label{fig:setup_keygen_comparison}
\end{figure}

\begin{figure}[h]
  \centerline{\includegraphics[width=1.2\columnwidth, trim={0cm 0cm 0cm 0cm}, clip]{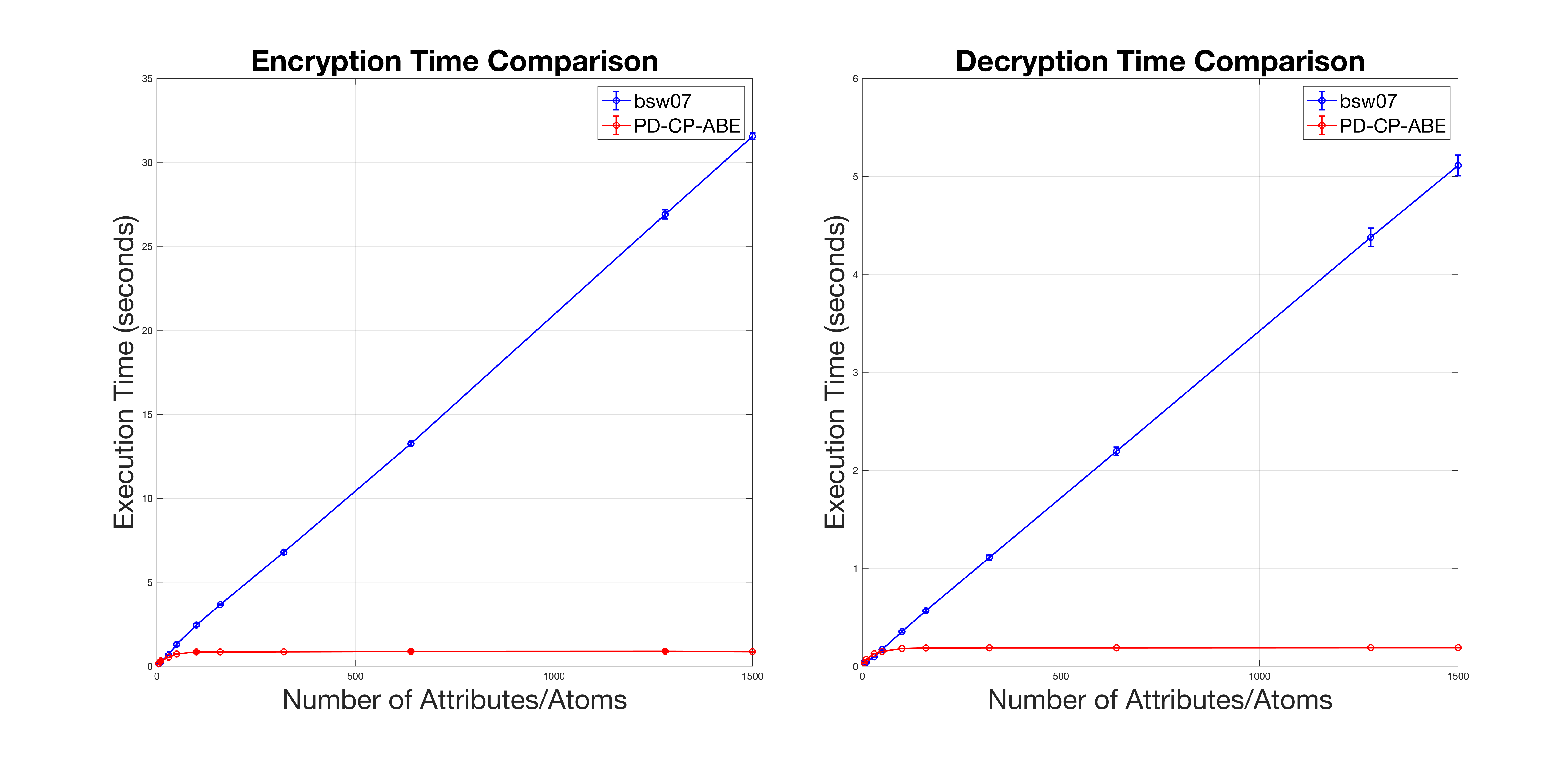}}
  \caption{Encryption Time and Decryption Time Comparison for bsw07 and PD-CP-ABE schemes.}
  \label{fig:encryption_decryption_comparison}
\end{figure}

% \begin{figure*}[htbp]
%     \centering
%     \begin{subfigure}[b]{0.45\linewidth}  
%         \centering
%         \includegraphics[width=0.7\linewidth, trim={0cm 0cm 0cm 0cm}, clip]{Plots/Setup_plot.png}
%         \caption{Setup time comparison.}
%         \label{fig:subfig1}
%     \end{subfigure}
%     \hfill
%     \begin{subfigure}[b]{0.45\linewidth} 
%         \centering
%         \includegraphics[width=0.7\linewidth, trim={0cm 0cm 0cm 0cm}, clip]{submission-template/Plots/keygen_plot.png} 
%         \caption{Keygen time comparison.}
%         \label{fig:subfig2}
%     \end{subfigure}

%     \vspace{0em} % Reduce vertical space between the rows

%     \begin{subfigure}[b]{0.45\linewidth}
%         \centering
%         \includegraphics[width=0.7\linewidth, trim={0cm 0cm 0cm 0cm}, clip]{submission-template/Plots/Encryption_plot.png}
%         \caption{Encryption time comparison.}
%         \label{fig:subfig3}
%     \end{subfigure}
%     \hfill
%     \begin{subfigure}[b]{0.45\linewidth}
%         \centering
%         \includegraphics[width=0.7\linewidth, trim={0cm 0cm 0cm 0cm}, clip]{submission-template/Plots/Decryption_plot.png}
%         \caption{Decryption time comparison.}
%         \label{fig:subfig4}
%     \end{subfigure}

%     \caption{Encryption scheme performance comparison.}
%     \label{Encryption Scheme performance comparison}
% \end{figure*}

\begin{table*}[htbp]
\centering
\caption{Comparative running times (in seconds) for PD-CP-ABE and bsw07 operations: Setup and Keygen.}
\label{tab:setup_keygen_performance}
\begin{tabular}{@{}lcccccc@{}}
\toprule
\textbf{Operation} & \textbf{Scheme} & \textbf{Number of Atoms} & \textbf{Avg Time (s)} & \textbf{Std Dev (s)} & \textbf{Median Time (s)} \\ 
\midrule
\textbf{Setup}  & bsw07     & -   & 0.0414   & 0.0005   & 0.0416   \\
                & PD-CP-ABE & -   & 0.0105   & 0.0001   & 0.0105   \\
\midrule
\textbf{Keygen} & bsw07     & 5    & 0.0258   & 0.0001   & 0.0258   \\
                &           & 10   & 0.0777   & 0.0285   & 0.0703   \\
                &           & 20   & 0.1035   & 0.0053   & 0.1015   \\
                &           & 30   & 0.1517   & 0.0077   & 0.1469   \\
                & PD-CP-ABE & 5    & 0.0299   & 0.0001   & 0.0299   \\
                &           & 10   & 0.0784   & 0.0284   & 0.0720   \\
                &           & 20   & 0.1285   & 0.0471   & 0.1007   \\
                &           & 30   & 0.1782   & 0.0555   & 0.1503   \\
\bottomrule
\end{tabular}
\label{Comparative running times (in seconds) for PD-CP-ABE and bsw07 operations: Setup and Keygen.}
\end{table*}

\begin{table*}[htbp]
\centering
\caption{Comparative running times (in seconds) for PD-CP-ABE and bsw07 operations: Encryption and Decryption.}
\label{tab:encryption_decryption_performance}
\begin{tabular}{@{}lcccccccc@{}}
\toprule
\textbf{Number of Atoms} & \multicolumn{4}{c}{\textbf{PD-CP-ABE}} & \multicolumn{4}{c}{\textbf{bsw07}} \\ 
\cmidrule(lr){2-5} \cmidrule(lr){6-9}
                         & \multicolumn{2}{c}{Encryption} & \multicolumn{2}{c}{Decryption} & \multicolumn{2}{c}{Encryption} & \multicolumn{2}{c}{Decryption} \\ 
\cmidrule(lr){2-3} \cmidrule(lr){4-5} \cmidrule(lr){6-7} \cmidrule(lr){8-9}
                         & Avg Time (s)  & Std Dev (s)   & Avg Time (s)  & Std Dev (s)   & Avg Time (s)  & Std Dev (s)   & Avg Time (s)  & Std Dev (s)   \\
5 atoms                  & 0.1652        & 0.0532        & 0.0358        & 0.0001        & 0.1513        & 0.0047        & 0.0396        & 0.0023        \\
10 atoms                 & 0.3167        & 0.0564        & 0.0702        & 0.0002        & 0.2495        & 0.0025        & 0.0397        & 0.0011        \\
30 atoms                 & 0.5423        & 0.0019        & 0.1291        & 0.0018        & 0.6890        & 0.0037        & 0.0982        & 0.0029        \\
50 atoms                 & 0.7298        & 0.0055        & 0.1509        & 0.0002        & 1.3068        & 0.1149        & 0.1706        & 0.0047        \\
100 atoms                & 0.8586        & 0.0843        & 0.1816        & 0.0002        & 2.4581        & 0.0928        & 0.3543        & 0.0101        \\
160 atoms                & 0.8574        & 0.0057        & 0.1876        & 0.0002        & 3.6726        & 0.0239        & 0.5675        & 0.0119        \\
320 atoms                & 0.8616        & 0.0168        & 0.1891        & 0.0004        & 6.7905        & 0.1239        & 1.1089        & 0.0248        \\
640 atoms                & 0.8879        & 0.0842        & 0.1894        & 0.0002        & 13.2531       & 0.1090        & 2.1940        & 0.0438        \\
1280 atoms               & 0.8926        & 0.0845        & 0.1903        & 0.0006        & 26.9057       & 0.2690        & 4.3785        & 0.0933        \\
1500 atoms               & 0.8719        & 0.0169        & 0.1902        & 0.0003        & 31.5530       & 0.1992        & 5.1119        & 0.1045        \\
\bottomrule
\end{tabular}
\label{Comparative running times (in seconds) for PD-CP-ABE and bsw07 operations: Encryption and Decryption.}
\end{table*}

The evaluation was conducted on a Debian server with the following configuration:
\begin{itemize}
    \item CPU: Intel(R) Xeon(R) CPU E5-2420 v2 @ 2.20GHz.
    \item Memory: 64 GB DDR3.
    \item Operating System: Debian GNU/Linux 12 (bookworm).
\end{itemize}

We compared our scheme against the foundational CP-ABE scheme proposed by Bethencourt, Sahai, and Waters \cite{Bethencourt2007} (referred to as bsw07 in the following section). The following functionalities were tested through practical runs:

\begin{itemize}
    \item Setup: A basic test to measure the initialization performance of the schemes.
    \item Keygen: Performance was tested using 5, 10, 20, and 30 attributes as parameters for key generation.
    \item Encryption/Decryption: Performance was tested using varying numbers of messages (5, 10, 30, 50, 100, 160, 320, 640, 1280, and 1500) as input. To simulate real-world scenarios, approximately and randomly 50 messages shared the same access policy. We use the word "atom" to represent each message for the following sections.
\end{itemize}

Each test was run 10 times, and the average result was calculated. The data obtained from these tests is presented in Table \ref{Comparative running times (in seconds) for PD-CP-ABE and bsw07 operations: Setup and Keygen.} and Table \ref{Comparative running times (in seconds) for PD-CP-ABE and bsw07 operations: Encryption and Decryption.}, while Fig.\ref{fig:setup_keygen_comparison} and \ref{fig:encryption_decryption_comparison} visually illustrate the performance comparison.

When comparing the two encryption schemes, bsw07 and PD-CP-ABE, we not only focus on the average execution time but also consider multiple aspects, including variance analysis, execution time stability, and scalability with different attributes/atom numbers.

\subsubsection{Setup Time Analysis} The PD-CP-ABE scheme significantly outperforms bsw07 in setup time, with an average execution time of 0.010532 seconds compared to 0.041434 seconds for bsw07. The standard deviation for PD-CP-ABE (0.000093 seconds) is considerably lower than that for bsw07 (0.000498 seconds), suggesting that PD-CP-ABE executes faster and more consistently and predictably.

\subsubsection{Key Generation (Keygen) Analysis} At 5 attributes, bsw07 exhibits a marginally faster average execution time (0.025796 seconds) compared to PD-CP-ABE (0.029930 seconds). As the complexity increases (20 and 30 attributes), bsw07 maintains a performance advantage with lower average execution times (0.103540 seconds and 0.151705 seconds, respectively) compared to PD-CP-ABE (0.128467 seconds and 0.178180 seconds).

\subsubsection{Encryption Performance} For smaller atom counts (5 and 10), bsw07 is slightly more efficient, with lower average encryption times. However, PD-CP-ABE becomes more efficient as the atom count increases (from 30 atoms onwards), offering significantly lower encryption times. The standard deviation for PD-CP-ABE is generally higher than for bsw07 at lower atom levels, indicating more variability in performance. However, as the number of atoms increases, the standard deviation of PD-CP-ABE remains relatively stable, indicating consistent performance despite the increased complexity.

\subsubsection{Decryption Performance} Across all atom levels, PD-CP-ABE consistently outperforms bsw07 in decryption time. The difference is particularly notable at higher atom levels (640, 1280, 1500), where PD-CP-ABE decrypts in approximately 0.19 seconds, while bsw07 takes significantly longer (up to 5.111895 seconds at 1500 atoms). The standard deviation for PD-CP-ABE remains extremely low across all atom levels, demonstrating that its decryption performance is faster and highly consistent.

\subsubsection{Formal Security Proof of PD-CP-ABE} We have used Tamarin Prover to analyse the security of the proposed encryption model. All security properties passed verification, and no counterexamples were found. This validates the robustness of our encryption scheme and confirms its security properties. The full details of security proof and instructions on how to replicate are available at the following link: \url{https://github.com/LimeFavoredOrange/PD-CP-ABE}. \\

In conclusion, the PD-CP-ABE scheme demonstrates clear advantages for applications that require CP-ABE encryption of multiple messages, such as digital wills. Not only does it offer efficient system setup, but it also provides superior performance in both encryption and decryption, particularly as the number of messages or the number of shared access policies increases. Although PD-CP-ABE exhibits relatively weaker performance in key generation compared to bsw07, the impact on overall scheme efficiency is minimal when weighed against its other performance benefits.

\section{Limitations}
\label{Discussion}
We take the first step towards developing an open-source PET solution that helps users achieve improved security and privacy for posthumous data management. While the system addresses essential challenges in proposing a working solution, certain limitations and potential enhancements warrant further exploration.

% \subsection{Digital Will File Format and Portability}
% \label{Digital Will File Format and Portability}
% While the design of the Digital Will File format has been conceptualized, its practical implementation and evaluation are left for future work. The current design proposes the use of XML as the foundation for the format, leveraging its structured and extensible nature to represent complex digital wills. However, several challenges remain to be addressed in ensuring the format’s portability across diverse services provider.

% Future research will focus on how to implement this file structure and realistically evaluate whether an XML-based digital will can be accurately parsed, extended, and redeployed by different providers without loss of data integrity or functionality.

%\subsection{Improving Access Policy Configuration}
%\label{Improving Access Policy Configuration}

\textbf{Improving Access Policy Configuration} Our system currently relies on users to manually set access policies for their data. As data volumes grow, this manual configuration becomes increasingly challenging. We suggest using artificial intelligence (AI) to automate the access policy configuration process to address this issue. For instance, users could define overarching principles and preferences, and the system accordingly applies appropriate tags and policies to each data item.

\textbf{Improving Server Trust Issues} We currently use the corresponding heir's key to encryption the data before distributing them. The reliance on server-side processing introduces potential vulnerabilities that could be exploited if the server is compromised. The need to generate encrypted versions of different users' keys also affects the practicality of the system. We aim to address this issue in future work by exploring more sophisticated cryptographic schemes. In addition, a dedicated protocol can be designed and developed that minimizes the trust in the server, ensuring that the confidentiality and integrity of the data are maintained even in adversarial environments.

\textbf{Future Work on Legal and Usability} The legal enforceability of digital wills across different jurisdictions remains an open challenge. Many countries require specific formalities, such as notarization, witnesses, or handwritten signatures, to make a will legally binding. Future iterations of this system will explore how to integrate legal requirements into the digital will process, ensuring compatibility with existing legal frameworks. This could involve working closely with legal experts to ensure that wills created and activated through the system can be verified and enforced in court. This solution also has various usability aspects that require further study and refinement.

\section{Conclusion}
\label{Conclusion}
This paper presents an open-source and Privacy Enhancing Technology (PET) for a third-party digital will management solution. `Beyond Life' securely and efficiently manages the posthumous distribution of users' digital assets at content-level granularity, leveraging a novel and efficient version of CP-ABE called PD-CP-ABE. We take a step forward to address technological gaps identified in the most recent studies (e.g., \cite{Reeves2024}) and devise a solution that addresses requirements specified for a working solution, including cross-platform support, decentralization, and portability. Leveraging blockchain technology and multi-cloud storage mechanisms, we bring transparency and control to address users' trust concerns when using a third-party solution for digital will management. The solution supports various deployment models, allowing users to customize their will's generation, storage, and processing (e.g., local or remote service). We have not studied legal and usability issues associated with `Beyond Life'. Further, we acknowledge the proposed solution required further technological refinements and extensions to optimise the user experience. From a security perspective, we believe there is a need to design solutions against various protocols and threat models in this domain before third-party digital will management solutions can be used at scale and in the real world. As the first open-source solution for digital will management, `Beyond Life' facilitates faster progress towards addressing gaps by helping research to be less theoretical and more pragmatic. 

\section{Project Source Code}
\label{Project Source Code}
The main system components are now uploaded to GitHub for public access (clickable Link below). The current public version uses pre-populated data since DWA connects to the user's private account.
\begin{itemize}
    \item \textbf{Digital Will App Repository:} \href{https://anonymous.4open.science/r/DigitalWillApp-9750/README.md}{\underline{Link}}

    \item \textbf{Digital Will App Home Page:} \href{https://expo.dev/preview/update?message=DigitalWillApp&updateRuntimeVersion=1.0.0&createdAt=2024-08-31T00%3A00%3A21.993Z&slug=exp&projectId=b773ff8b-02ea-4b8e-831a-8cc85f0d6c89&group=665411e2-4ddb-462b-b14b-eb25f7275677}{\underline{Link}}

    \item \textbf{PD-CP-ABE Repository:} \href{https://github.com/LimeFavoredOrange/PD-CP-ABE}{\underline{Link}}

    \item \textbf{PD-CP-ABE endpoint Postman Collection:} \href{https://app.getpostman.com/run-collection/23135719-b685e954-bcd1-43e5-95f0-b3038725ad3f?action=collection%2Ffork&source=rip_markdown&collection-url=entityId%3D23135719-b685e954-bcd1-43e5-95f0-b3038725ad3f%26entityType%3Dcollection%26workspaceId%3D1f0a1601-87f1-4963-a3f1-9194ade5e62b}{\underline{Link}}

    \item \textbf{PD-CP-ABE Docker Image:} \href{https://hub.docker.com/r/xinzhang9091/charm-crypto-ubuntu22.04}{\underline{Link}}

\end{itemize}

\appendices

\section{Data Availability}
\label{Data Availability}
The artifacts necessary to reproduce the work presented in this paper, including implementing the core system components, such as the encryption scheme, Digital Will Broker, and the Document Decomposition/Assembly modules, are made publicly available under an open-source license on GitHub. The user interface application DWA will be publicly downloadable on personal mobile devices. The project uses synthetic data, such as sample wills and access policies, and no sensitive or personal data is involved. Documentation detailing the setup, configuration, and usage will also be provided to ensure reproducibility. In cases where sharing specific materials is restricted due to privacy or legal concerns, appropriate justifications will be offered.

\section{Flow diagram of system interaction}
\label{Appendix B}

\begin{figure}[h]
  \centering
  \includegraphics[width=\linewidth]{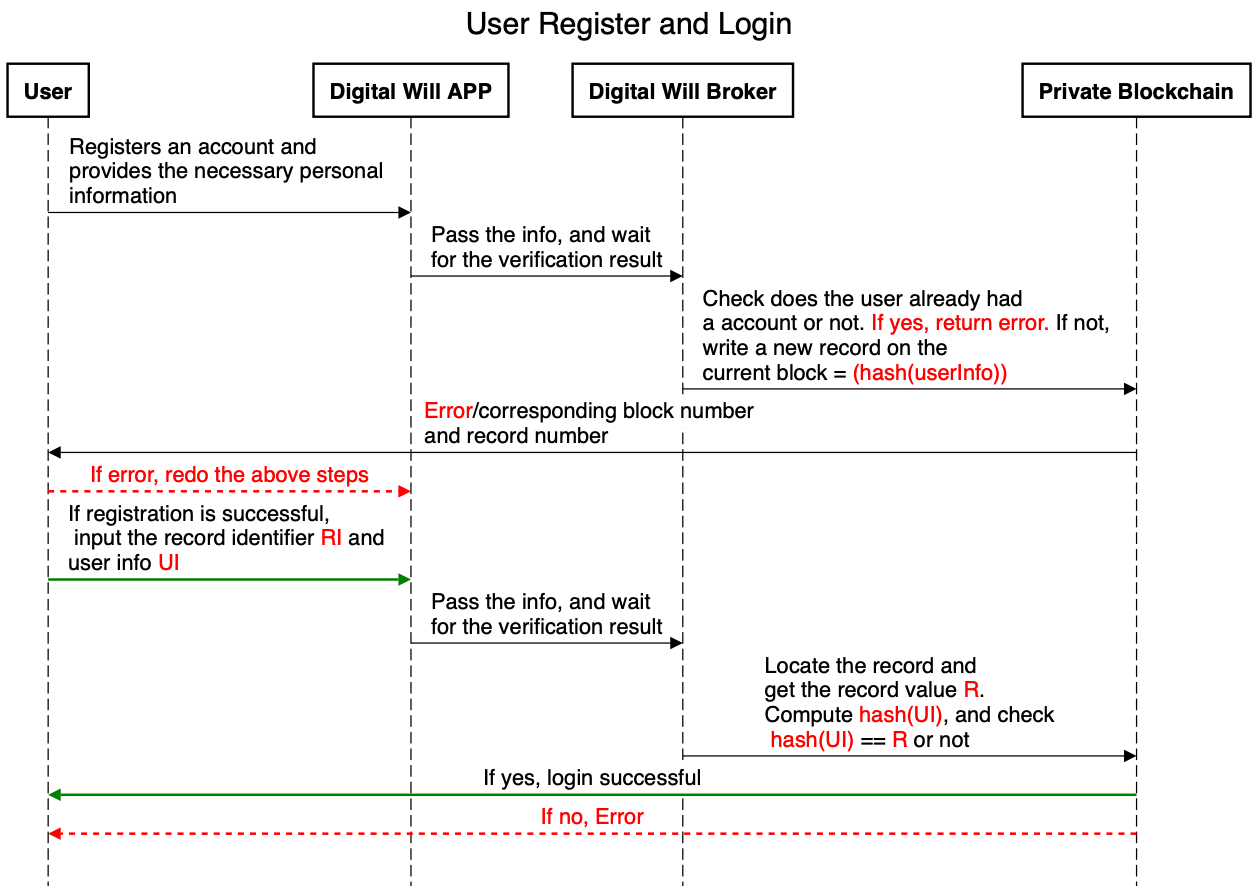}
  \caption{The interaction flow for user register and login}
\end{figure}

\begin{figure}[h]
  \centering
  \includegraphics[width=\linewidth]{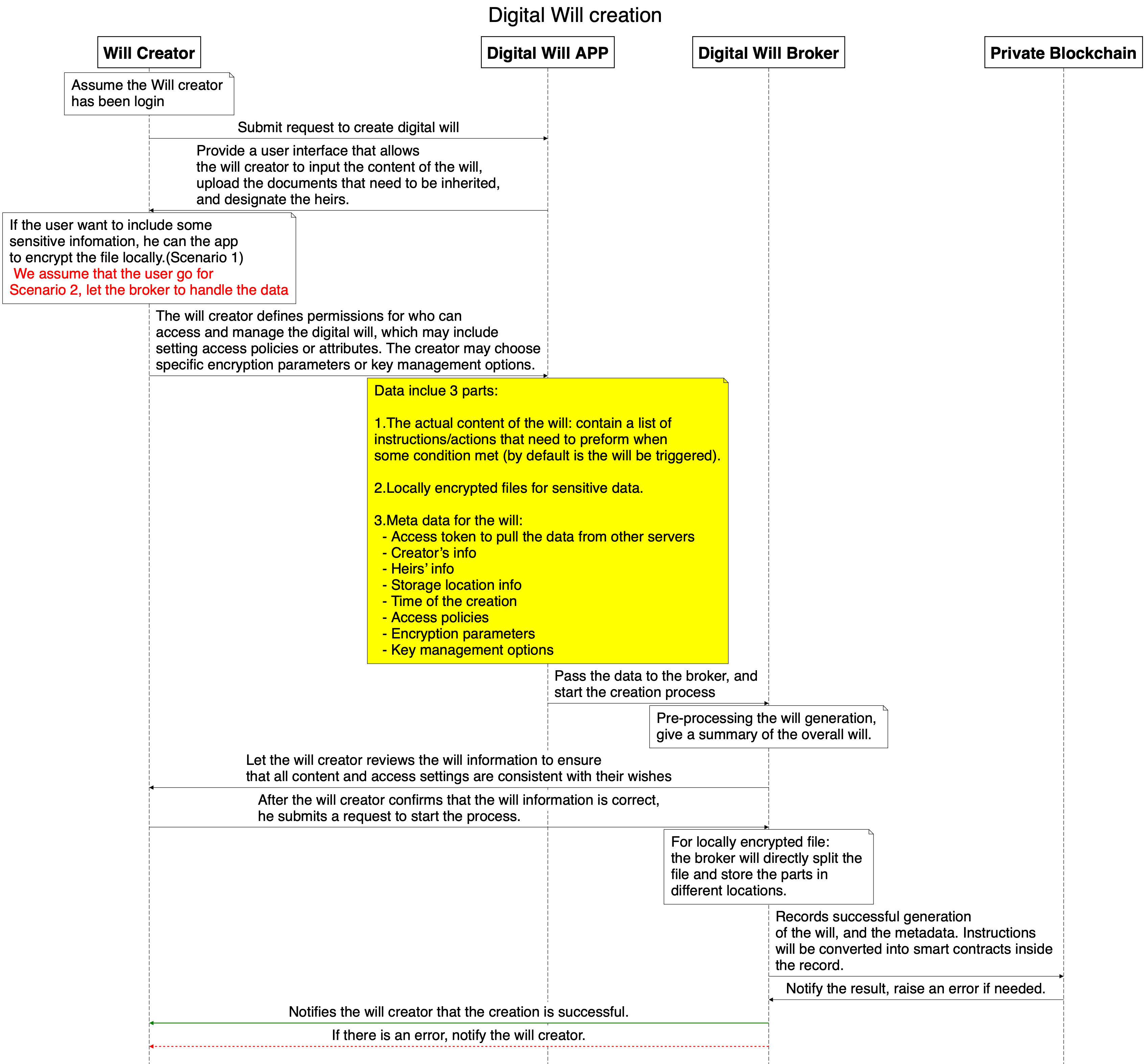}
  \caption{The interaction flow for Digital Will Creation}
\end{figure}

\begin{figure}[h]
  \centering
  \includegraphics[width=\linewidth]{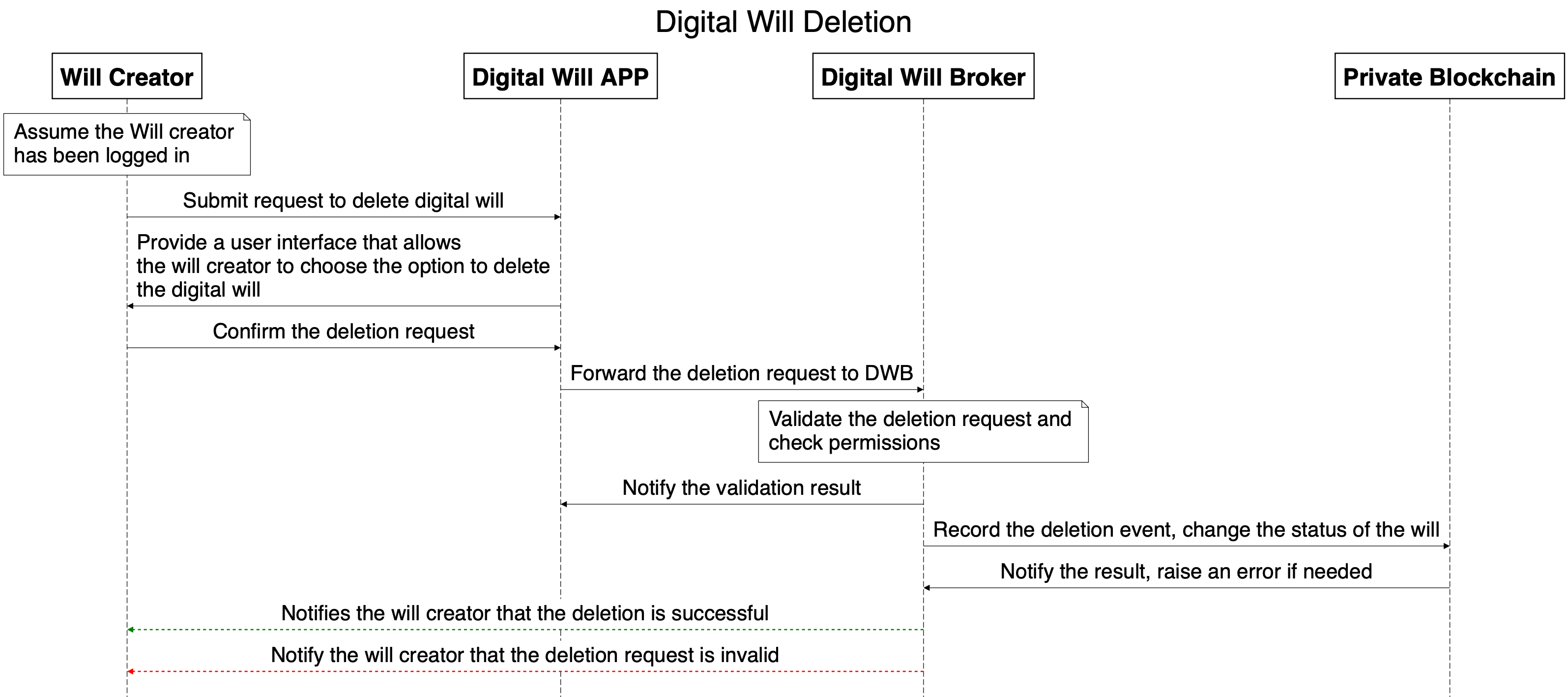}
  \caption{The interaction flow for Digital Will Deletion}
\end{figure}

\begin{figure}[h]
  \centering
  \includegraphics[width=\linewidth]{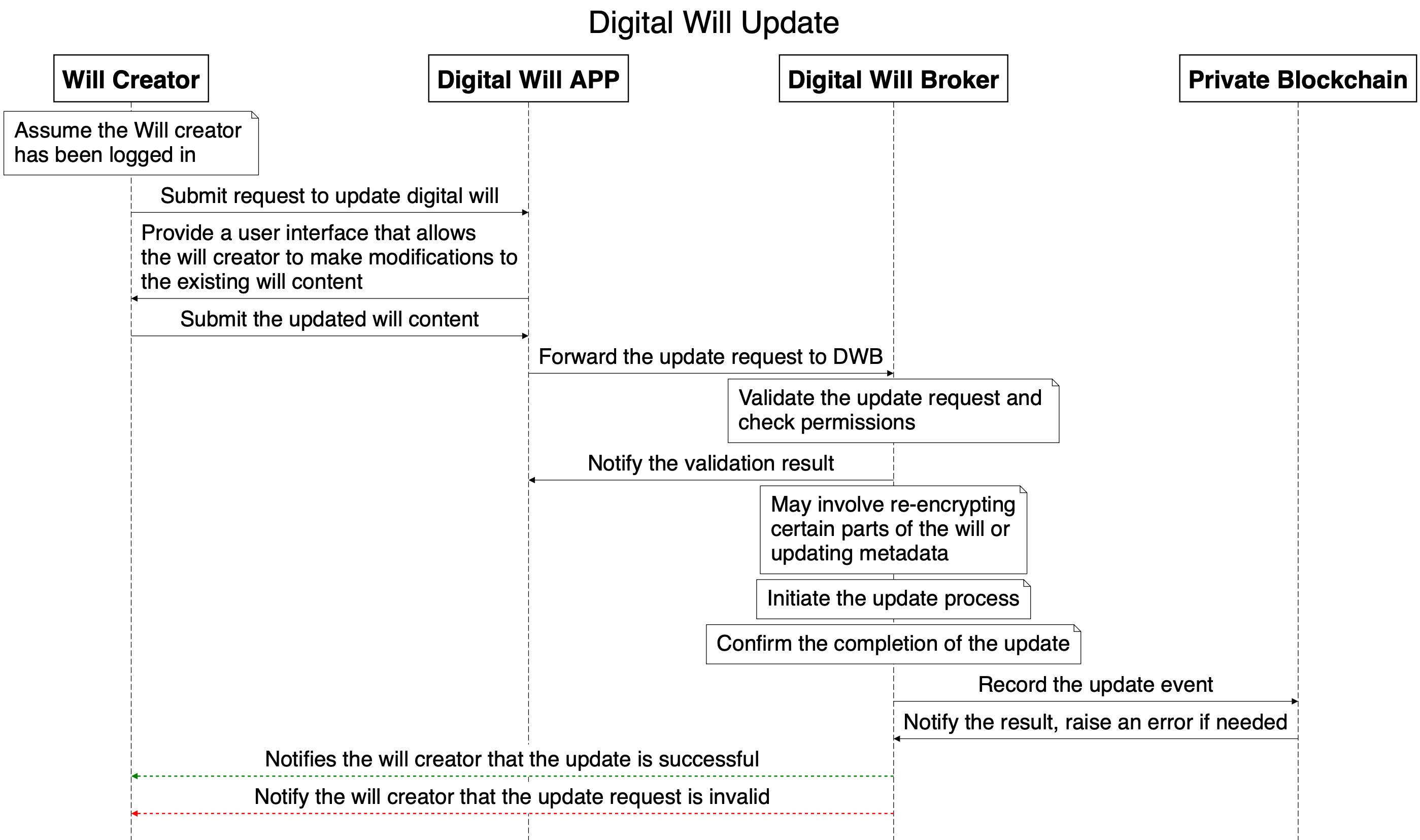}
  \caption{The interaction flow for Digital Will Update}
\end{figure}

\begin{figure}[h]
  \centering
  \includegraphics[width=\linewidth]{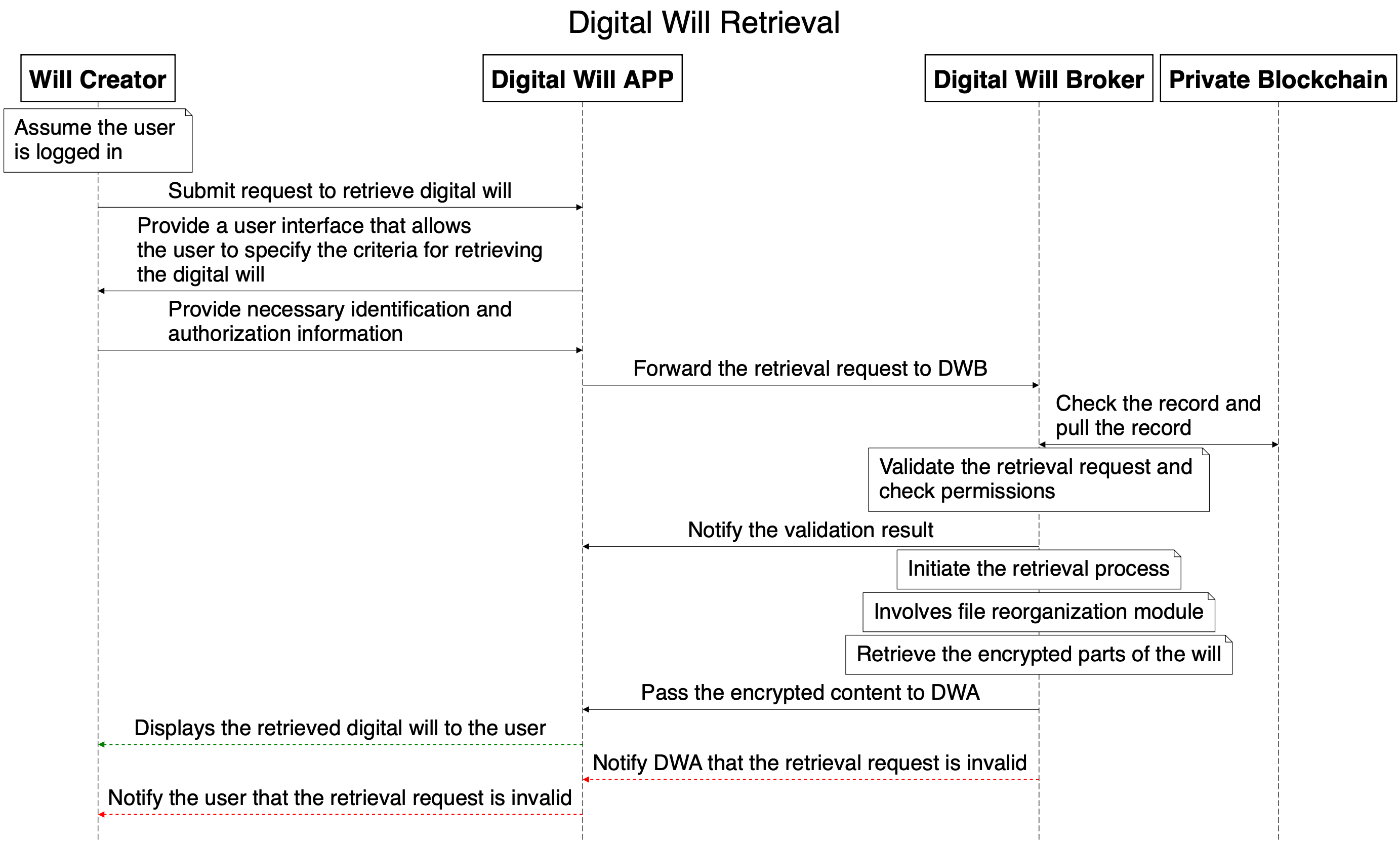}
  \caption{The interaction flow for Digital Will Retrieval}
\end{figure}

\begin{figure}[h]
  \centering
  \includegraphics[width=\linewidth]{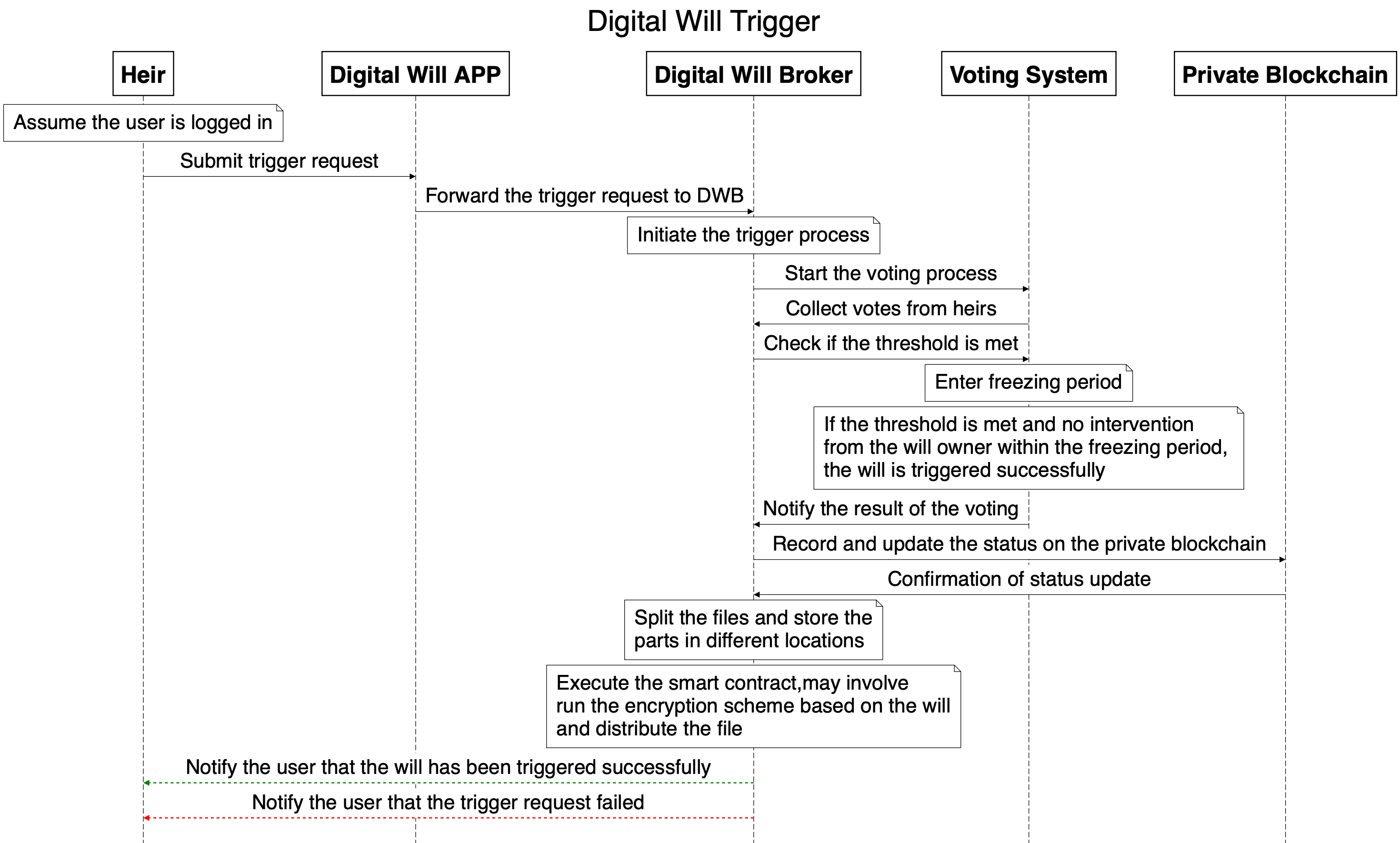}
  \caption{The interaction flow for Digital Will Trigger}
\end{figure}

\section{Screenshots of the Digital Will App}
\label{Appendix C}

\begin{figure}[h]
  \centering
  \includegraphics[width=\linewidth]{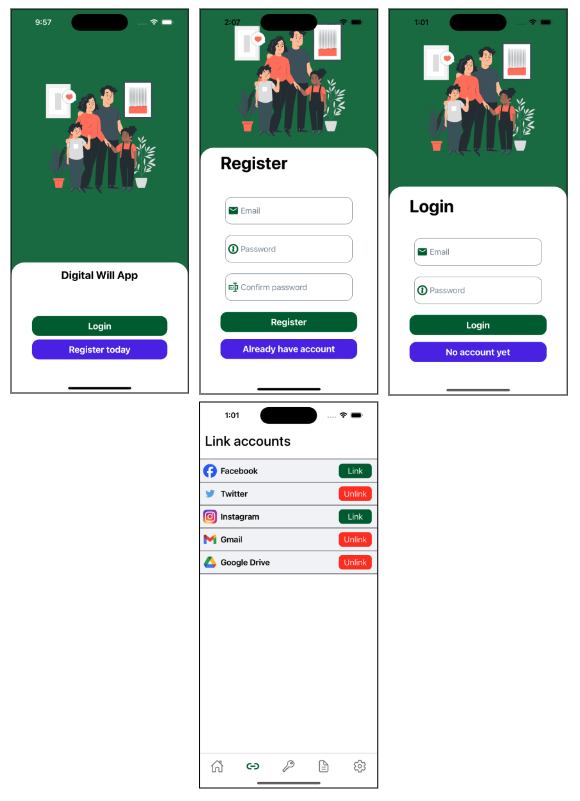}
  \caption{Screenshots of the Beyond Life App showing the landing, registration, login, and account linking interfaces}
\end{figure}

\begin{figure}[h]
  \centering
  \includegraphics[width=\linewidth]{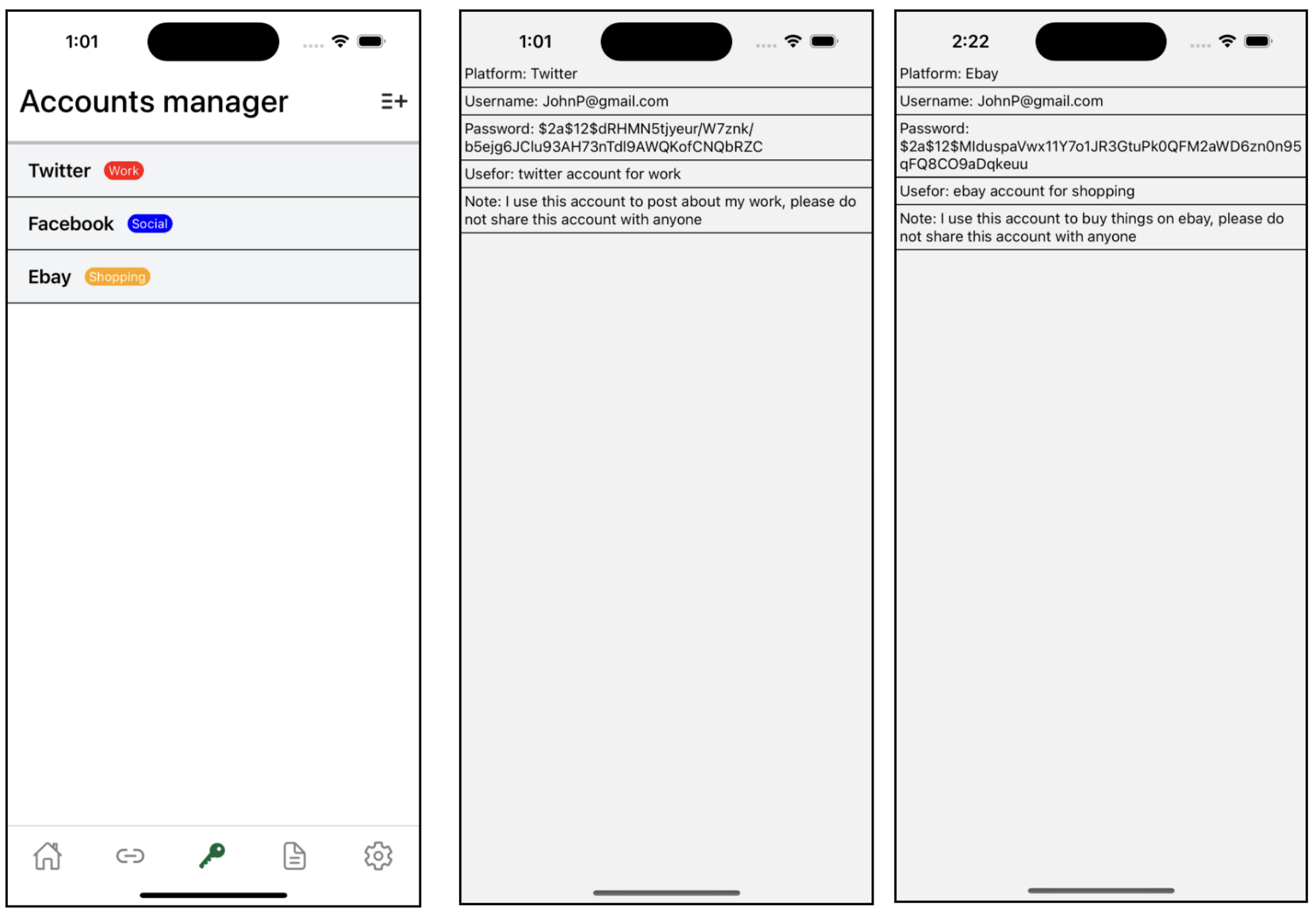}
  \caption{Screenshots of the Accounts Manager feature in the Beyond Life App}
\end{figure}

\begin{figure}[h]
  \centering
  \includegraphics[width=\linewidth]{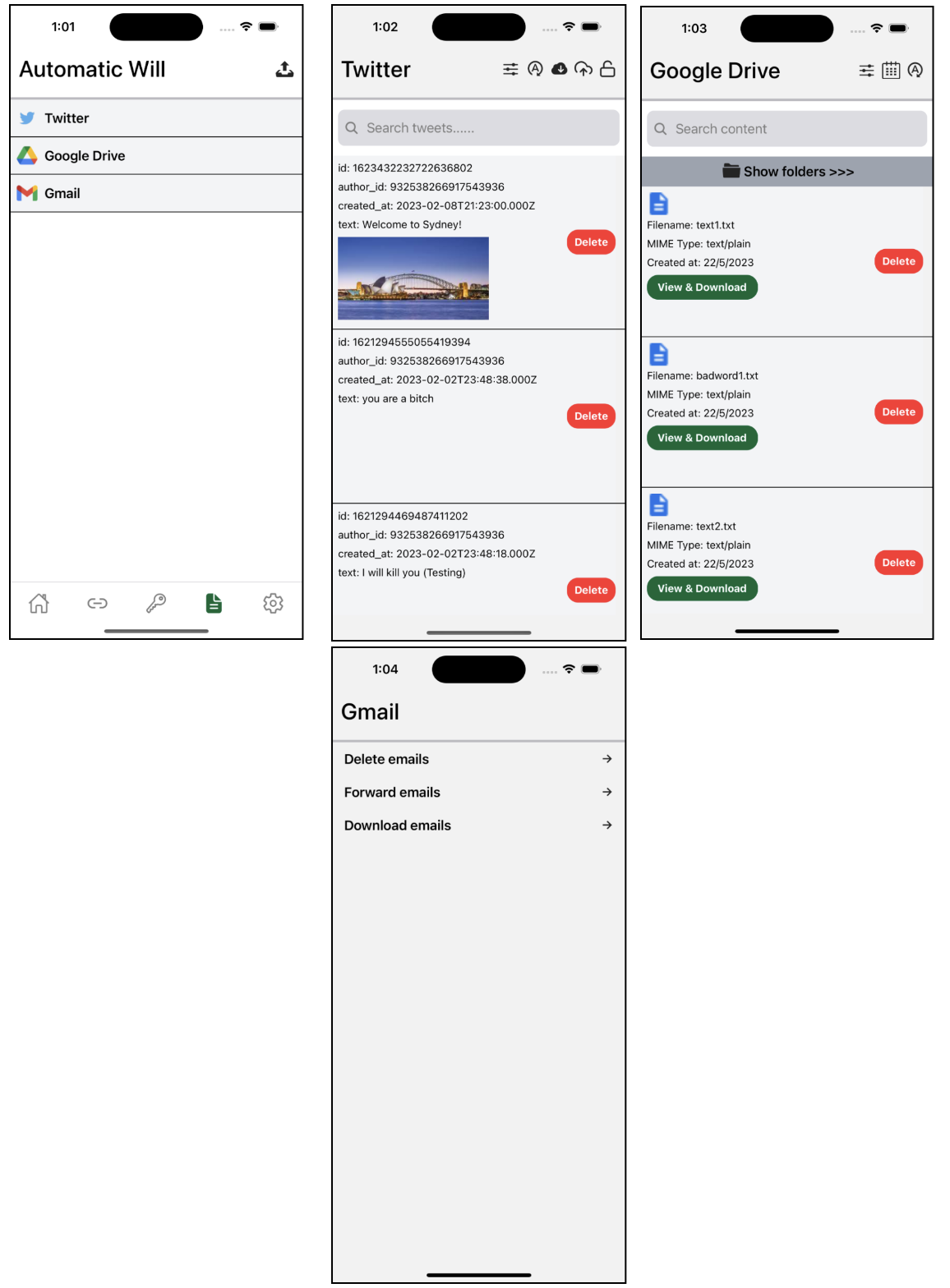}
  \caption{creenshots of the Automatic Will feature in the Beyond Life App}
\end{figure}

\begin{figure}[h]
  \centering
  \includegraphics[width=\linewidth]{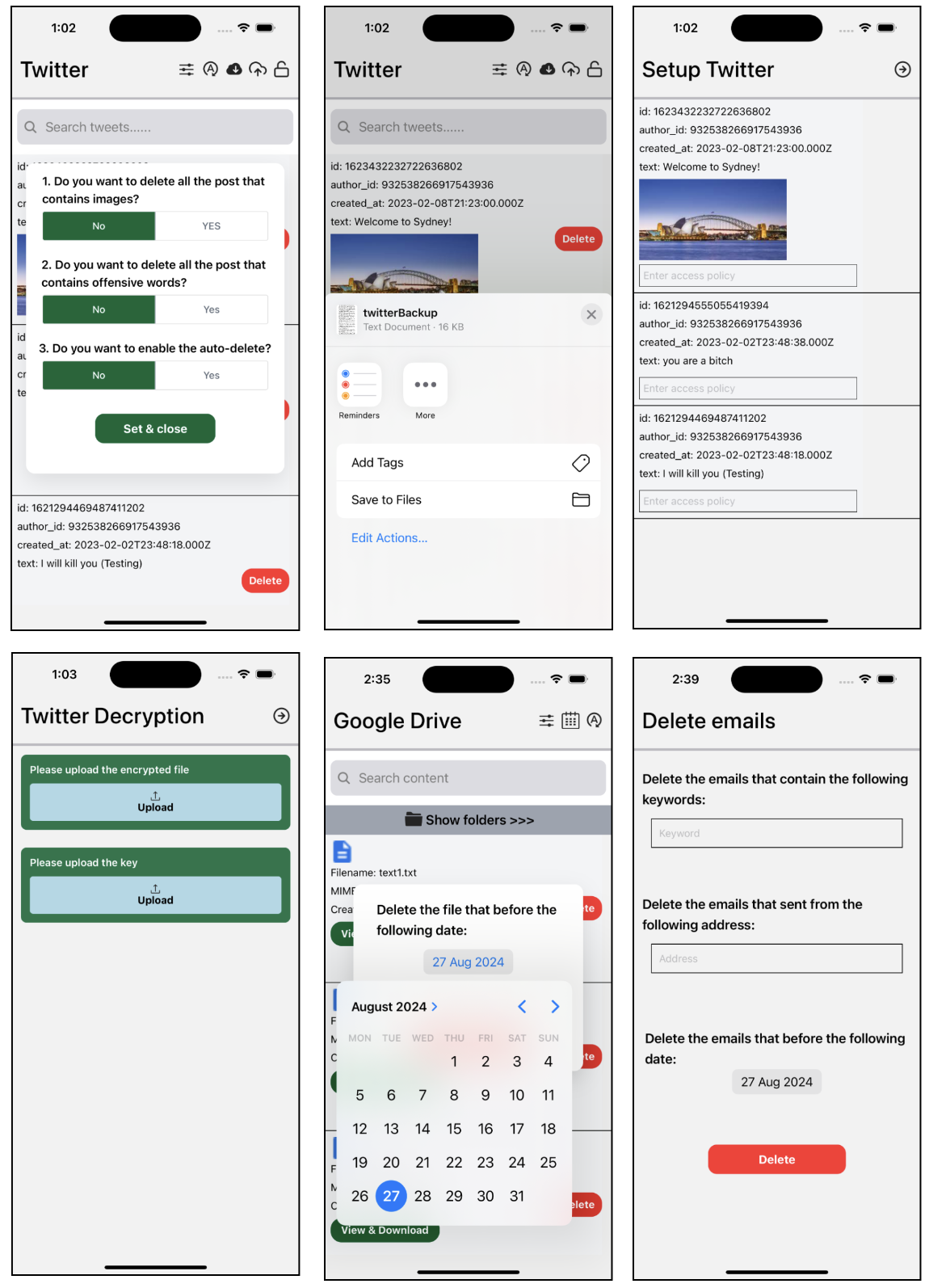}
  \caption{Screenshots of the will settings and management features in the Beyond Life App}
\end{figure}

\begin{figure}[h]
  \centering
  \includegraphics[width=0.6\linewidth]{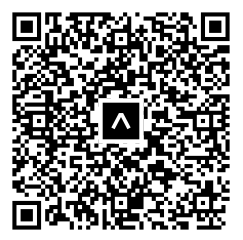}
  \caption{QR code of Digital Will App preview page}
\end{figure}

\end{document}